# Comparative transcriptome analysis reveals key epigenetic targets in SARS-CoV-2 infection

Marisol Salgado-Albarrán[1,2,†], Erick I. Navarro-Delgado[3,†], Aylin Del Moral-Morales[1,†], Nicolas Alcaraz[4,5], Jan Baumbach[2], Rodrigo González-Barrios[3*] and Ernesto Soto-Reyes[1*]

[1] Departamento de Ciencias Naturales, Universidad Autónoma Metropolitana-Cuajimalpa (UAM-C), Mexico City, Mexico.

[2] Chair of Experimental Bioinformatics, TUM School of Life Sciences Weihenstephan, Technical University of Munich, Munich, Germany.

[3] Unidad de Investigación Biomédica en Cáncer, Instituto Nacional de Cancerología-Instituto de Investigaciones Biomédicas, Mexico City, Mexico.

[4] The Bioinformatics Centre, Department of Biology, University of Copenhagen, Copenhagen 2200.

[5] National Institute of Genomic Medicine, Periférico sur 4809, Arenal Tepepan, 14610 Mexico City, Mexico.

* To whom correspondence should be addressed. Tel: +52(55)58146500 ext. 3879; Email:esotoreyes@cua.uam.mx. Correspondence may also be addressed to Tel:(55)56280400 ext. 70036; Email: rodrigop@ciencias.unam.mx

[†] These authors contributed equally to this work

## ABSTRACT

COVID-19 is an infection caused by SARS-CoV-2 (Severe Acute Respiratory Syndrome coronavirus 2), which has caused a global outbreak. Current research efforts are focused on the understanding of the molecular mechanisms involved in SARS-CoV-2 infection in order to propose drug-based therapeutic options. Transcriptional changes due to epigenetic regulation are key host cell responses to viral infection and have been studied in SARS-CoV and MERS-CoV; however, such changes are not fully described for SARS-CoV-2. In this study, we analyzed multiple transcriptomes obtained from cell lines infected with MERS-CoV, SARS-CoV and SARS-CoV-2, and from COVID-19 patient-derived samples. Using integrative analyses of gene co-expression networks and de-novo pathway enrichment, we characterize different gene modules and protein pathways enriched with Transcription Factors or Epifactors relevant for SARS-CoV-2 infection. We identified EP300, MOV10, RELA and TRIM25 as top candidates, and more than 60 additional proteins involved in the epigenetic response during viral infection that have therapeutic potential. Our results show that targeting the epigenetic machinery could be a feasible alternative to treat COVID-19.

**BACKGROUND**

The coronavirus family (CoV) are non-segmented, positive-sense and enveloped RNA viruses that have been identified as the cause of multiple enteric and respiratory diseases in both animals and humans [1]. Three major CoV strains of this family have caused recent human pandemics: Middle East respiratory syndrome coronavirus (MERS-CoV) in 2002-2003 [2], severe acute respiratory syndrome coronavirus 1 (SARS-CoV) in 2012 and SARS-CoV-2 in 2020 [3]. The most recent one was identified in Wuhan, China by the end of 2019 and is the etiological origin of an atypical pneumonia known as COVID-19, which has caused a global outbreak and is one of the top sixth public health emergencies of international concern (Lai et al., 2020) with 38,269,885 confirmed cases and 1,088,515 deaths as of October, 2020, leading to the biggest CoV pandemic in modern times [4].

By being intracellular pathogens, viruses' infection strategy requires the continuous subordination and exploitation of cellular transcriptional machinery and metabolism in order to ensure its own expansion. To do so, the host genome expression must be used and, to be successful, this will depend on chromatin dynamics and transcription regulation, which are principally ruled by epigenetic mechanisms, such as DNA methylation, histone post translational modifications (HPTM), and transcription factors (TF) [5]. During a viral infection, it has been reported that epigenetic and transcriptional changes occur for both sides: the infected cell promotes an antiviral environmental response, leading to the induction of pathways to survive, while the virus switches off the expression of critical anti-viral host cell genes [6,7].

Several studies have reported the importance of epigenetic modifications in viral infections. In influenza virus, specific gene promoter methylation [8], decreased H3K4me3 (a hallmark of active chromatin) [9], histone acetylation in H3 and H4 histones, and increased levels of H4K20me2 and unmodified H3K36 and H4K79 have been reported [10]. Interestingly, these HPTM do not always trigger the same mechanisms and lead to similar phenotypes; for example, depletion of H3K79me2, an epigenetic mark that is usually increased upon viral infections due to an upregulation of DOT1L, results in impaired viral growth in human cytomegalovirus infection [11], while enhancing the replication in influenza virus [10]. However, these mechanisms usually lead to host transcriptional inactivation, which contribute to the altered cellular transcription produced by viral infections.

Regarding CoVs, few experimental studies have been conducted to unravel the epigenetic proteins and marks involved in their infection and pathogenesis in MERS-CoV and SARS-CoV, being specially scarce in SARS-CoV-2 due to its recent appearance. For MERS-CoV and SARS-CoV, different outcomes have been reported, such as the mechanisms used to control the interferon-stimulated genes, which involves H3K27 methylation in MERS-CoV but not in SARS-CoV [12], and the ones used to down-regulate antigen-presenting molecules, which involves DNA methylation in MERS-CoV and not in SARS-CoV [8]. These studies show that epigenetic mechanisms are highly important in the host gene expression control carried out by the virus and that, despite the phylogenetic closeness, these mechanisms can be very different between strains, highlighting the need to understand the epigenetic processes that play a role in SARS-CoV-2 infection.

Integrative computational methods are promising approaches used to generate research hypotheses, generate consensus regulatory networks and to describe deregulated processes in SARS-CoV-2 infection [13,14]. Nevertheless, they have overlooked key epigenetic and transcription factors that underlie the infected phenotype. Since drugs that target the epigenetic landscape of diseased cells have shown a great potential and have proved to be game-changing as complementary treatments of complex diseases, such as cancer [15], the identification of these key epigenetic and transcription factors becomes highly important in our current context, where popular regimen candidates for treating COVID-19, such as Remdesivir, Hydroxychloroquine, Lopinavir and Interferon have shown to have little or no effect on reducing mortality of hospitalized COVID-19 individuals [16].

In this work, we gathered publicly available RNA-seq data from SARS-CoV-2, SARS-CoV and MERS-CoV infected cell lines and patient samples, and performeddifferential expression analyses together with weighted gene co-expression network analysis to identify unique and shared central epigenetic players in SARS-CoV-2, SARS-CoV and MERS-CoV. Candidate genes were further prioritized by integrating DEGs enrichment tests, gene-coexpression network and viral-host protein-protein interaction network analysis to propose potential key epigenetic proteins involved in SARS-CoV-2 infection. Finally, we identified currently approved drugs that target key epigenetic drivers of SARS-CoV-2 infection, and thus they are potential new therapeutic approaches for COVID-19.

**METHODS**

**Data processing and differential expression analysis**

Raw sequencing data was trimmed with Trimmomatic version 0.39 [17] using the parameters ILLUMINACLIP 2:30:10 LEADING:3 TRAILING:3 SLIDINGWINDOW:4:15 MINLEN:36; and the quality of reads was evaluated with FastQC version 0.11.9 [18]. Technical replicates (when existing) were merged and each biological replicate was aligned to the GRCh38 v33 human genome with STAR version 2.7.3 [19] using the mapping parameters suggested in Jin et al. [20]: (--outFilterMultimapNmax 100 --winAnchorMultimapNmax 100). To estimate the abundance of the transcripts accounting for coding and non-coding genes as well as repetitive elements, we used TETranscripts version 2.1.4 [20] with the multi mode. Raw count tables were used for differential expression analysis using DESeq2 [21]. Differentially expressed genes (DEGs) were identified with a p adj. < 0.05 and abs($\log_2$ fold change) > $\log_2(1.5)$.

**Viral transcripts quantification**

Viral transcriptome was constructed with the 11 gene sequences reported in SARS-CoV-2 genome (NCBI Reference Sequence NC_045512.2). Viral transcript expression was quantified in each trimmed RNA-seq file of SARS-CoV-2 infected samples with Salmon v 1.3.0 [22].

**Virus and patient DEGs**

Virus-associated gene sets were obtained with the intersection of DEGs identified in all the cell lines infected with the corresponding virus, except for SARS-CoV-2. For SARS-CoV, the intersection between the cell lines infected consisted of 182 genes (SARS-CoV-DEGs); for MERS-CoV the intersection was 1139 genes (MERS-CoV-DEGs); and for SARS-CoV-2, the intersection between at least 3 out of the 4 cell lines was used instead and consisted in 909 genes (SARS-CoV-2-DEGs) (Supplementary Table 2). Patient-associated gene set was obtained with the DEGs in at least 2 out of the 3 conditions (543 genes, patient-DEGs) (Supplementary Table 2).

**Epigenes catalogue**

To build the Epigenes catalogue, 4 different databases were used: EpiFactors [23], Histome [24], dbEM [25] and the manually curated TF list from Lambert et al. [26]. TFs' functional annotation was taken from Lambert et al. [26]. The final list consisted of 2161 genes (776 epifactors, 1348 TFs and 41 categorized as both TF and epifactor).

**Co-expression analysis**

Count matrices of the analyzed cell lines were filtered to remove low-expressed genes using the function filterByExpr from edgeR [27] while accounting for the treatment (i.e. virus infection) and cell type in the filtering design. Following, normalization of gene counts was performed with vst function from DESeq2[21] (treatment and cell type of each sample were included into the design matrix and accounted for these effects with the blind argument). The gene co-expression network was built with the $\log_2$ fold changes of each biological sample compared with the controls of the same biological condition by applying the formula (1).

$$(1)\ \log_2 FC_i = \log_2(SC_i / ACC_i)$$

Where SC and ACC correspond to the normalized counts of gene *i* in the infected and controls samples respectively. The resulting matrix containing the log2FoldChanges per sample was used to construct the weighted gene co-expression network with the WGCNA package [28]. A soft threshold of 9 was used to construct the network and modules were identified with a minimum size of 20. Modules whose expression was similar were merged using a dissimilarity threshold of 0.25, resulting in a total of 24 modules. Finally, the module-eigengene pearson correlation of each module with the viruses was tested.

**Enrichment analysis**

Gene Ontology enrichment analyses were performed using clusterProfiler [29] in virus associated and patient gene sets. For the differential expression analyses of infected cell lines, the enrichment of GO terms in DEGs was tested using the expressed genes on each particular comparison as background. For the co-expression network, the enrichment of GO terms was tested in each module using the genes of the full network as background.

Epigenes, virus-associated DEGs and TF-target enrichment analyses were performed with gProfiler2 [30] using a custom gmt file or the TRANSFAC database included in the package for TF-target enrichment. The correction method used was g:SCS and an adjusted p-value significance threshold of 0.05. As background, all the genes annotated in the co-expression network were used for epigenes and TF-target enrichment and the expressed genes in each virus for virus-associated DEG enrichment.

**Virus-host network construction**

Virus-human interactions were obtained from Gordon et al. [31] and Stukalov et al. [32]. The human protein-protein interaction network (PPI) was obtained from IID version 2018-11 [33] using only the experimentally validated interactions ("exp", "exp;ortho", "exp;ortho;pred" or "exp;pred"). After homogenizing the viral protein nomenclature, the three sources of interactions were merged to create the entire virus-human PPI, followed by removal of duplicated edges and self loops. The final integrated network contained 32 viral nodes, 17524 human nodes and 329054 edges. The mapping of viral transcript counts to viral proteins in the PPI was based on the reference sequence annotation (NCBI Reference Sequence NC_045512.2) and the data provided in Supplementary Data from Gordon et al. 2020 [31].

**Epigene selection**

For modules 4, 6, 8, 10, 11 and 12, relevant epigenes were selected based on whether they satisfied at least one of the following criteria: (1) its shortest path length with viral proteins, (2) the correlation value between its expression and the expression of viral proteins and (3) its module membership (MM) value, a measure of the correlation between a gene expression profile and the module eigengene, which is highly related to the intramodular connectivity, and gene significance (GS) the correlation of a gene with an external trait (viral infection) [34].

(1) The shortest path length was calculated between all pairs of viral proteins and human proteins in the PPI network with the igraph package version 1.0.0 [35]. The retained epigenes were the ones whose shortest path length with at least one viral protein was less than 3.

(2) Pearson's correlation coefficient was computed between the count values of viral transcripts and count values of epigenes in infected cell lines. Epigenes with p value < 0.05 and abs(correlation_estimate) > 0.5 with at least one viral transcript were selected.

(3) Epigenes with abs(MM) > 0.8 in the corresponding module of the co-expression network were retained.

For modules 1 and 7, epigenes with abs(MM) > 0.8 and abs(GS) > 0.3 were selected.

*De novo* **pathway enrichment**

*De novo* pathway enrichment analysis for modules 4, 6, 8, 10, 11 and 12 was performed with KeyPathwayMiner [36], the built virus-human PPI network, the full list of viral proteins as positive nodes and a customized input indicator matrix for each module containing as active genes those which belonged to any of the following categories: (1) it was a SARS-CoV-2 DEG, (2) it was a patient DEG or (3) it was an epigene selected as described above. The parameters used for all the analyses were the Greedy search algorithm, INES search strategy, remove border exception nodes, L=0, and K=0 for modules 4 and 12, K=2 for module 6, and K=3 for modules 8, 10 and 11.

**Drug identification**

All approved and non-approved drugs targeting the genes/proteins contained in each network were obtained with CoVex [37] by mapping the gene names to uniprot IDs, using the closeness centrality algorithm and the following parameters: result size= 50000, disabled hub penalty, disabled max degree, include indirect drugs=FALSE and include non-approved drugs=TRUE.

**DATA AVAILABILITY**

Raw RNA-seq data was obtained from the Sequence Read Archive (www.ncbi.nlm.nih.gov/sra) of the National Center for Biotechnology Information (NCBI), U.S. National Library of Medicine, and the Genome Sequence Archive in BIG Data Center (bigd.big.ac.cn/), Beijing Institute of Genomics (BIG), Chinese Academy of Sciences (Supplementary Table 1).

**RESULTS**

**SARS-CoV-2, SARS-CoV and MERS-CoV induce different transcriptional and epigenetic responses during infection in pulmonary cell lines**

In order to identify the genes that change their expression in pulmonary cell lines (Calu-3, MRC-5, A549 and NHBE) due to infection of Coronaviruses such as MERS-CoV, SARS-CoV or SARS-CoV-2, differential expression analysis was performed in RNA-seq data.

As a first approach, we evaluated the overlapping DEGs identified for each virus regardless of the cell type and in common among viruses. For MERS-CoV and SARS-CoV, the overlap among all cell conditions was considered for SARS-CoV-2, the overlap among 3 out of the 4 cell conditions was used, since the NHBE cell line showed a small number of DEGs most likely because these cells are derived from normal bronchial epithelial cells [38,39] (Supplementary Figure 1A). We observed that the majority of the virus-associated genes are unique for each virus and a small proportion is shared among them. Specifically, only 3 genes were differentially expressed during infection in cell lines regardless of the virus evaluated (Figure 1A). Furthermore, Gene Ontology (GO) enrichment analysis (Figure 1B) shows that the top 10 enriched GO terms are different for each virus, except "cellular response to lipopolysaccharide", shared between SARS-CoV-2 and SARS-CoV; however, the three viruses share terms related to immune response processes (Supplementary Figure 1B). The latter shows that, despite their phylogenetic relationship, the main changes in gene expression driven by MERS-CoV, SARS-CoV-2 and SARS-CoV infection are divergent at both levels: at the DEGs and the cellular processes, suggesting that each virus uses specific molecular strategies during infection.

Subsequently, we inspected the DEGs with epigenetic or transcriptional regulatory function present among viruses, hereinafter referred as epigenes. A comparative analysis of the differentially expressed epigenes among the three viruses revealed that only *INO80D*, a regulatory component of the chromatin remodeling INO80 complex, is shared among them. MERS-CoV and SARS-CoV only share the histone deacetylase *HDAC9*; while MERS-CoV and SARS-CoV-2 share *DUSP1*, *KDM6B*, *CHD2* and *GADD45A*. Between SARS-CoV and SARS-CoV-2, we found *PBK MYSM1*, *ZNF711* and *PCGF5* (Figure 1C, Supplementary Figure 1C). In addition, given that TFs are also key elements in gene remodeling and regulation, we evaluated the ones differentially expressed across viruses and none of them was affected in all conditions. However, MERS-CoV and SARS-CoV share *ZNF484* and *CEBPD*; SARS-CoV and SARS-CoV-2 share *ZEB1*, *ZEBTB20*, *NR4A1* and *FOXN2,* and between MERS-CoV and SARS-CoV-2 15 shared TFs were found, including *RELB*, *JUN*, *FOSB*, *E2F8*, among others (Figure 1C, Supplementary Figure 1D).

Furthermore, the analysis showed that the differentially expressed epifactors belong to a wide range of functional categories, such as histone writers, histone readers, histone erases, Polycomb group proteins, chromatin remodeling, DNA modifications, among others (Figure 1D). In addition,regarding differentially expressed TFs, cell lines infected with SARS-CoV-2 show differential expression of TFs of the STAT (mediators of the cellular response to cytokine) and IRF (interferon-regulatory factor) family, which are not differentially expressed in MERS-CoV and SARS-CoV (Figure 1E). We noted that most of the TFs that are differentially expressed and shared between two or more of the Coronaviruses infected cells are members of Znf transcription factor family (ZNF436, 448, 543, 597, 773, XSCAN12, ZEB1, ZDTB20, KLF10 and HIVEP1) bHLH family (MXD1 and MXD4), involved in CCAAT/Enhancer Binding Protein (C/EBP) (DDIT3 and CEPPD), NF-κB complex (RELB), AP-1 complex (FOSB, JUN), ETS family (SPDEF) and E2F transcription factor, among others (Supplementary Figure 1C).

Lastly, we evaluated the changes in genes expression of repeat elements after viral infection (Supplementary Table 3). We found that 47, 22 and 319 repeat elements are differentially expressed in SARS-CoV-2, SARS-CoV and MERS-CoV infected cell lines, respectively. In SARS-CoV-2, the repeat elements belong predominantly to the LINE (13 elements), LTR (17 elements) and DNA repeat elements (17 elements) families. Similarly, for SARS-CoV and MERS-CoV we found LINE (5 and 59 elements), LTR (10 and 179 elements) and DNA repeat (5 and 65 elements). Interestingly, SINE elements are not differentially expressed (only 3 elements found in MERS-CoV) and few Satellite elements were identified (3, 1, and 10 in SARS-CoV-2, SARS-CoV and MERS-CoV, respectively) The elements L1MA4:L1:LINE, L1PA8A:L1:LINE and LTR54:ERV1:LTR are shared among all viruses. Notably,the L1 or LINE-1 elements are the only autonomous transposons that remain active in the human genome, and are mainly repressed by epigenetic mechanisms such as HPTMs (H3K9me3). The latter, along with the fact that SARS-CoV-2 infected cells overexpress the histone demethylases KDM7A and KDM6B that target the heterochromatin histone marks such as H3K9me and H3K27me [40,41] (Supplementary Figure 1C), suggest that SARS-CoV-2 infection could promote an open chromatin conformation, thus affecting the transcriptional expression and the derepression of the repeated sequences.

**SARS-CoV-2 transcriptional effect is tissue-dependent in COVID-19 patient-derived samples**

Afterwards, we evaluated the transcriptional response in patient samples infected with SARS-CoV-2 to assess their resemblance to the previously observed results in cell lines. For this purpose, we obtained datasets from samples of bronchoalveolar lavage fluid (BALF), lung and peripheral blood mononuclear cells (PBMC). From the differential expression analysis, we found 16 DEGs shared among all three sample types (Figure 2A), the majority of them with a different fold change direction in PBMC compared to LUNG and BALF, and 6 genes with a consistent fold change direction (Figure 2B). This is most likely explained by the higher cell-lineage similarity between the Lung and BALF samples than with PBMC, which could lead to a more similar transcriptional response to the infection. In addition, GO enrichment analysis shows that most of the DEGs were related to the immune response to viral infection such as leukocyte migration and humoral immune response (Figure 2C). Following, we evaluated the similarity of these results with the data observed for SARS-CoV-2 in infected cell lines by comparing the overlap between the virus-associated genes with the DEGs present in at least 2 out of the 3 sample types in patients (hereinafter referred as patient-DEGs, 543 genes). We found 61, 15 and 33 genes in common with SARS-CoV-2, MERS-CoV and SARS-CoV, respectively. In particular for SARS-CoV-2 and patients, 6 TFs (*STAT5A, MAFF, IRF9, MXD1, FOSB* and *STAT4)* and no epifactors were identified. We also found, at a lesser extent, shared DEG between samples of SARS-CoV-2 infected patients and MERS-CoV and SARS-CoV infected cell lines, which are likely to be non-specific viral-responding immune genes (Figure 2D). Finally, we found that 804, 1 and 20 repeat elements are differentially expressed in BALF, PBMC and LUNG samples, respectively, being LTR elements the most differentially expressed in all samples (Supplementary Table 3). Collectively, these results show that the gene expression changes promoted by SARS-CoV-2 infection in patients are diverse among tissue types; however, despite these differences, immune response processes are the main processes affected across all sample types.

**SARS-CoV-2 and MERS-CoV infection induce different transcriptional fold changes that involve the same gene co-expression modules, which recapitulate the expression profiles in COVID-19 patient-derived samples**

So far, our analyses showed that cell lines and patient samples infected with SARS-CoV-2 exhibited DEGs related to immunological processes, which has been previously described by Blanco-Melo et al. [42] which is congruent with our results. However, differential expression analysis often overlooks the subtle differences in several genes that altogether can be responsible for major changes in global transcriptional regulation. Weighted gene co-expression network analysis overcomes this limitation by studying the expression of thousands of genes in the same analysis [28]. Thus, we expanded our previous results by including a co-expression analysis to identify gene modules associated with each viral infection and genes that play central roles within them.

We constructed the co-expression network with the log2 fold changes of each sample compared to its controls. After identifying the modules, we calculated the correlation between each module and the different traits, where we found that out of the 24 total modules identified, 13 were significantly correlated to the infection of any of the three viruses (Supplementary Figure 2). Specific modules are associated with MERS-CoV (module 1), SARS-CoV (module 7), and SARS-CoV-2 (module 9 and 10) (Figure 3); shared modules were also identified. Notably, more than half of the modules (7 out of 13) are significantly associated with both SARS-CoV-2 and MERS-CoV; contrary to SARS-CoV-2 and SARS-CoV, which have only one module jointly associated. In addition, when a module is associated with both MERS-CoV and SARS-CoV-2, the correlation sign is the opposite between them. The latter indicates that the transcriptional responses induced by MERS-CoV and SARS-CoV-2 are more similar to each other than with SARS-CoV because both involve the same co-expression modules (and consequently the same genes and biological processes), and that the response of the deregulated genes appears to be the different.

Further analysis with GO enrichment analysis (Supplementary Figure 3), shows that genes in modules 11 and 12 are involved in the host cell response to viral infection, while modules 4, 5, 7, 8, 10 and 13 were associated with intracellular processes used by the viruses during the infection, such as DNA replication, translation, ribosome biogenesis and protein folding. Interestingly, module 6 was found related to epigenetic processes, particularly, transcriptional activation of promoters. These results show that, in addition to the immunological processes identified in our previous differential expression analysis, with the co-expression network we find that the transcriptional response to SARS-CoV-2 and MERS-CoV infection contain genes that also participate in RNA translation, DNA replication and epigenetic regulation.

Following, in order to determine the modules that might be more relevant to SARS-CoV, MERS-CoV, or SARS-CoV-2 infection in terms of epigenetic regulation, we conducted enrichment analyses to identify an over-representation of epigenes, virus-associated gene sets, DEGs found in patients and targets of TFs (Figure 3, SupplementaryTable 4).

When integrating these analyses, we found module 1 relevant for MERS-CoV and module 7 for SARS-CoV infection since they are exclusive for these viruses according to the co-expression analysis (Figure 3). To determine the most important epigenes for each of these modules, we evaluated the eigengene-based connectivity (Module Membership, MM) to find hub genes, and the gene significance (GS) of each gene. Subsequently, we prioritized them by identifying drugs targeting them. After looking for candidate drugs that targeted these central players, we found approved medication for CDK7 and PCNA for SARS-CoV and NCOA1, NR1H2, PRKAB2, CLOCK, KDM1B and ATF2 (Supplementary figure 4).

Regarding SARS-CoV-2, module 9 was uniquely associated with it, but also other modules standed out. First, module 4 had a consistent behaviour between cell lines and patients, since we found it negatively correlated to SARS-CoV-2, while being enriched in downregulated genes in BALF and LUNG samples; in addition, it was enriched in epifactors, suggesting an important role in viral-related epigenetic modifications carried out by these genes. A similar phenomenon is observed for module 12, which was positively correlated with SARS-CoV-2 infection and enriched with upregulated genes in patients and with SARS-CoV-2 DEGs. Additionally, modules 10 and 11 were positively correlated to SARS-CoV-2 and enriched in SARS-CoV-2 DEGs, with module 11 being also enriched with patient's DEGs. Module 6 was negatively associated with SARS-CoV-2 in the co-expression network and enriched with epigenes, and module 8 was enriched with upregulated genes in PBMC and negatively associated with SARS-CoV-2 in the co-expression network, while being enriched with epifactors and SARS-CoV-2-DEGs (Figure 3). Finally, the enrichment of TFs targets in each module was evaluated to identify the ones that could explain the co-expression patterns of the genes within the module. With this analysis, it was found that Module 4 is enriched in the target genes of the transcriptional factors MTA1, MORC2, and RBM34 that belong to the same module (Supplementary Table 4), being MTA1 and RBMC34 differentially expressed in BALF samples. It is worth to mention that in most of these modules epigenes showed a higher MM than the rest of the genes (Module 1: W = 164249, $p < 0.05$, Module 2: W = 1497, $p < 0.05$, Module 4: W =

51768, p < 0.05, Module 6: W = 1359, p < 0.05, Module 7: W = 3741, p < 0.05, Module 12: W = 113820, *p* < 0.05, Module 13: W = 2182049, *p* < 0.05), evincing their central role within their modules.

Collectively, these results show that the transcriptional response to infection of SARS-CoV-2 and MERS-CoV is more similar than the one observed in SARS-CoV. Furthemore, both viruses involve primarily the same gene modules but with a different extent of transcriptional change in host cells during infection, which extends our previous observations in the differential expression analysis. Importantly, the virus-correlated co-expression modules either recapitulate the changes in gene expression observed in different COVID-19 patient sample types or are enriched with epifactors, and also contain genes involved in several biological processes related to viral infection, suggesting that the data obtained in the cell lines could recapitulate what was found in infected patients

**Protein-protein interaction network analysis provides additional therapeutic alternatives and new targets for drug development for COVID-19**

To prioritize epigenes that play a key function in each co-expression module relevant for SARS-CoV-2 (modules 4, 6, 8, 10, 11 and 12), we examined them at the protein-protein interaction (PPI) level in the context of SARS-CoV-2 infection. We constructed a PPI network containing all experimentally validated human protein interactions [33] and the reported virus-host protein interactions from Gordon et al. 2020 and Stukalov et al 2020 [31,32]. Using the virus-human PPI network, we performed *de novo* pathway enrichment analysis with KeyPathwayMiner [36] to extract the largest network using a selection of epigenes as input, while also taking into account the SARS-CoV-2-DEGs and patient-DEGs previously identified. The selection of epigenes for each module was based on their shortest path length with viral proteins, their expression correlation with viral genes and their MM.

All genes contained in the networks identified (Figure 4) provide insights about the molecular machinery involved in SARS-CoV-2 infection, since the genes are either differentially expressed in infected cell lines or patients, or they are hub-epigenes in the co-expression analysis.

For module 4, the network obtained contains mainly epifactors. Notably, DNMT1 directly interacts with the viral protein ORF8 and with TRIM28, to which it is also highly co-expressed. Other relevant epigenes in the networks are SIRT6 (highly co-expressed and interactor of TRIM28), SENP3, MTA1 (a TF whose targets are also enriched in module 4; Supplementary Table 4), BAP1 (differentially expressed in patients) and HCFC1. Furthermore, MEPCE, a snRNA methyl phosphate capping enzyme, is differentially expressed in patients and interacts with viral protein NSP8. Module 6 contains BRD4, which directly interacts with E viral protein and is highly co-expressed with EP300, a histone acetyltransferase. Other relevant epigenes are GMEB2 (differentially expressed in MERS-CoV) and SETD1B (related the trimethylation of histone H3 at Lys 4, H3K4me$^3$, a unique epigenetic histone mark related to transcriptional activation), which interacts with TRIM28, present in module 4 [43]. For module 8, notable epigenes are CENPF, differentially expressed in SARS-CoV and SARS-CoV-2 and directly interacting with NSP13; and TOP2A, differentially expressed in SARS-CoV infected cell lines and patients. In module 10, RELA (also known as nuclear factor NF-κB p65 subunit) and TRIM25 are exception nodes that interact with the TFs SMAD3, ZNF277 and UBE2D3. Also, viral proteins ORF7B and ORF3 interact with FXYD2, STEAP1B and TMEM156, which are differentially expressed in SARS-CoV-2 infected cell lines. For module 11, IRF7 (interferon regulatory factor 7) and STAT5A interact with EP300. Module 12, also contains epigenes of interest, such as MOV10 (Putative helicase MOV-10) that interacts with the N protein, TRIM25, RELA and TLE1, which has a direct interaction with viral protein NSP13. Further genes which are classified as hub-epigenes and are also differentially expressed in cell lines or patients are FOS, CEBPD, NR4A1, PRDM1, PCGF5, ZNF652, IRF2, ZEB2 and SP110.

Briefly, we identified relevant TFs known to participate in Coronavirus infection and support the veracity of our results, such as TFs from the STAT family (STAT1, STAT2, STAT5A), interferon regulatory factors (IRF7, IRF2), cytokines (CCL3, CCL4, CXCL1 and CXCL8), and FOS and JUN, members of the AP-1 complex [44]. However, we also identify important genes that appear to be drivers of SARS-CoV-2 infection; such as the epifactors MOV10 and EP300, the TF RELA and the enzyme TRIM25, since they are exception connectors (genes that do not belong to the specific module, but are important in the protein pathway found) in more than one module (modules 6, 10 and 11), and belong to modules enriched in genes that participate in histone H3-K4 methylation and in the response to interferon gamma. EP300 is a histone acetyltransferase that was also identified in SARS-CoV-2 infected cell lines [13]. Additionally, we found that MOV10, a putative helicase, also participates in SARS-CoV-2 infection. The TF RELA has been increasingly recognized as a crucial modulator of the response to SARS-CoV-2 infection [13,45] and is part of the NF-κB complex, along with RELB [46], which is differentially expressed in MERS-CoV and SARS-CoV-2 infected cell lines (Supplementary Figure 1D). TRIM25 is a ubiquitin ligase required for production of INF-1 and is inhibited by Nucleocapsid of SARS-CoV [47]. Finally TRIM28 (also known as KAP1) has been shown to interfere with viral integration into host genome [48] and represses the expression of repeat elements of the LINE family, in particular L1NA4 [41], which was previously identified as differentially expressed in cell lines infected with the three Coronaviruses.

Afterwards, we evaluated whether the proteins in the networks had annotated drugs targeting them. We found drugs for 69 out of the total 260 proteins, being PLAU, RELA, NEK6, NR1H4, PTGS2, PRKDC, ESR1, NR3C2, TTK, TOP2A, ADRB2, HDAC4, TRIM25, STK10, RPS6KA5 and EP300 the ones with the most drugs identified (more than 20). A total of 799 drugs were found, where Erlotinib, Imatinib, Lapatinib, Sunitinib, S-adenosyl-L-homocysteine, Quercetin, Tandutinib, RAF-265, Pictilisib, Neratinib and Fedratinib are the drugs with more targets (more than 5; Supplementary Table 5). Relevant epigenes that have associated medication are shown in Table 1.

Most notably, RELA is targeted by SC-236, Bortezomib, Indoprofen (an anti-inflammatory) and Betulinic Acid, whose derivatives show anti-HIV activity [49,50]. EP300 is targeted by curcumin, a molecule with anti-inflammatory properties [51]. The latter proposes RELA and EP300 as new potential drug target candidates for SARS-CoV-2 infection, not only because they participate in immune-related processes, but also because they belong to the cellular epigenetic machinery used by the virus during infection. Furthermore, self-evident immune-related targets STAT5A, STAT1, CXCL8, CCL3, JUN and FOX are also good candidates for treatment. Finally, the proteins MOV10, TRIM25 and TRIM28 do not have an associated drug, thus they are good candidates for drug development, as well as other relevant epigenes shown in Table 2.

Together, network analysis at the protein level allowed the identification of several epigenes that are part of the molecular machinery used by the virus during infection. Epigenes that participate in immune response through different mechanisms (response to interferon or NF-kB complex) are among the main genes identified and are evident drug target candidates for COVID-19 because they already have associated drugs targeting them (such as STAT5A, CXCL8 and CCL3). Furthermore, new candidate druggable epigenes were also identified, notable examples are EP300 and RELA, which are targeted by drugs with anti-inflammatory or antiviral properties; and TRIM25, TRIM28 and MOV10, which are good candidates for drug development.

**DISCUSSION**

Cells are in constant adaptation with their environment, in fact they can sense and respond to different stimuli by changing their transcriptional patterns. This cellular plasticity allows cells to adapt almost immediately to insults, including virus infections [52]. Epigenetic proteins and TFs are one of the main elements involved in the transcriptional response of cells during viral infection. These elements can be used as protein targets for drug identification and treatment. In this work we aimed to identify key TFs and proteins involved in the epigenetic response to viral infection of SARS-CoV-2, SARS-CoV and MERS-CoV by integrating co-expression and de *novo* pathway enrichment analyses. One of our main findings is that the transcriptional response (regarding DEGs and significantly co-expressed modules) induced by SARS-CoV-2 infection is more similar to MERS-CoV than to SARS-CoV. However, unique modules, patterns and DEG were found in each CoV. Despite they belong to the coronavirus family, each one has unique characteristics that could influence its pathogenicity and virulence. This finding agrees with a recent study that has found specific biological process deregulations in SARS-CoV-2 infected cell lines, which are not found in other CoVs [14]. In addition, different transcriptional change patterns have been observed between MERS-CoV and SARS-CoV during the infection; these changes are not recapitulated by phylogenetic relationships since, in some groups of genes, MERS-CoV-infected transcriptional behaviour appears to be more similar to the more remotely related influenza H5N1 virus infection [12].

Furthermore, the contrasting transcriptional response induced by the infection of SARS-CoV-2 and MERS-CoV in several modules suggests that genes in those modules participate in both viral infections but with a different mechanism, which leads to distinct pathways of infection that could explain the dissimilar phenotypes observed in both diseases. Divergent fold change trends, such as the ones described in this study, have been previously described in MERS-CoV and SARS-CoV infections to limit the host type I interferon (IFN-I) response, where predominant active and repressive epigenetic marks in involved genes are the opposite between both CoVs [12]. In our study, we present a list of epigenes and biological processes whose fold change trend is the opposite between MERS-CoV and SARS-CoV-2; further investigation on them could shed light on the mechanisms responsible for the differences in pathogenesis and outcome of both viral infections.

We further identify at the protein interaction level, that several TFs take part mainly in the immunological response to viral infection. One example is NF-κB, whose p65 subunit (also known as *RELA*) is a central part in the protein interaction network for SARS-CoV-2. NF-κB induces the expression of several pro-inflammatory cytokines, including IL-6, CCL2 and CCL3 [53], which had been found in high levels in COVID19 patients [54]. On the other hand, TRIM25, an ubiquitinase, is essential for the activation of NF-κB and the production of IL-6 [55]. TRIM25 is over-expressed in cell lines infected with SARS-CoV-2 but not in those with MERS-CoV, which furthermore suggest that NF-κB could be a medullary part of the host immune response against SARS-CoV-2. The previous observation is reinforced by the fact that it was observed that RELA directly interacts with histone acetyltransferase EP300, and both proteins interact with various components of the AP-1 complex such as FOS, JUND, and FOSL1. AP-1, EP300 and NF-κB regulate chromatin accessibility in the proximal promoter region of IL-6 and CCL2, both pro-inflammatory cytokines [56,57]. The p300/CBP complex is one of the best characterized cofactors of NF-κB and specifically binds RELA and acetylates it along with the surrounding histones [46]. It is known that adults older than 65 years have higher NF-κB levels compared to younger adults [58] and some authors had suggested that this may be one reason older adults are more susceptible to develop the severe form of COVID-19 [59].

According to our results, SARS-CoV-2 infection modifies the expression of several TFs of the interferon regulatory factor (IRF) and STAT families, which are primarily involved in the immune response against pathogens. STAT1 and STAT2 are key elements of the signaling induced by type I interferons, this proteins form a dimer upon interferon mediated phosphorylation and, together with IRF9, form the complex ISGF3 that activates the transcription of interferon stimulated genes [60]. Our results also showed that IRF9 is upregulated in cell lines infected with SARS-CoV-2; however, module 12's interactome showed that STAT2 and STAT1 interact with IRF2. IRF2 is a negative regulator of IFNα and its inhibition causes an increase in the antiviral response induced by IFNα [61]. This fact further suggests an impairment of interferon type I stimulated genes activation, as previously described as a hallmark of SAR-CoV-2 infection [62]. On the other hand, IRF1 and IRF7 were also upregulated in SARS-CoV-2 infected cell lines. IRF7 is a key TF for IFNα expression, and it has been previously identified as a hub gene for SARS-CoV-2 infection together with IFR9 and STAT1 [63]. It is also interesting that IRF7 lost of function mutations were associated with severe COVID-19 patients [64] and with the development of life-threatening influenza in children [65] which suggest that inhibition of IRF7 activity is crucial for SARS-CoV-2 pathology.

Viruses have been reported to use epigenetic machinery to take advantage of the cell and hijack its regulatory capacity for their own benefit [12]. The epigenetic machinery can be affected by coronaviruses in this same sense, and this can happen either by promoting alterations in the epigenetic code, such as DNA methylation and post-translational modifications of histones, or directly by promoting the dysregulation of enzymes and other proteins associated with the epigenome.

We found that among the deregulated epifactors with histone acetylation function are HDAC9 and SIRT1 enzymes. In this sense, it has recently been reported that the SIRT1 protein (a class 3 HDAC) was positively regulated in the lung of patients with severe COVID-19 comorbidities [66]. Likewise, another work demonstrated that under conditions of cellular energy stress, SIRT1 can epigenetically regulate the ACE2 receptor [67]. Also, it has been observed that treatment with non-steroidal anti-inflammatory drugs can inhibit SIRT1 activity, which in turn could affect ACE2 expression [68]. Accordingly, it has been postulated that some diseases such as lupus where the epigenetic dysregulation is implicit in this disease could facilitate the entrance of SARS-CoV-2 into the host cells [69].

Interestingly, the enzymes HAT1, HDAC2 and KDM5B have been reported to also potentially regulate ACE2 in human lungs. KDM5B has gained interest, because it is associated with other viral infections such as the hepatitis B virus [70], and potentially with SARS-CoV-2 [66]. Remarkably, in breast cancer cells, it has been shown that inhibition of this enzyme triggers a robust interferon response that results in resistance to infection by DNA and RNA viruses [71]. In this regard, we observed several deregulated KDMs in the different coronavirus infections, in which KDM6B stands out by being deregulated in both MERS-CoV and SARS-CoV2 infection. KDM6B is a specific demethylase of H3K27me3, which acts as a repressive histone mark. Although it remains to be fully studied, it is associated with the regulation of a wide range of genes involved in inflammatory agents, development, cancer, viral infection response, senescence and is an important host response against environmental, cellular stress [72]. Therefore, adding to the above, it is suggested that demethylases, such as KDM6B, are potential epigenes that are affected during SARS-CoV-2 infection and can be presented as potential targets for the treatment of COVID-19. However, this should be further studied.

Several epigenes previously involved in response to viral infections stood out in our protein interaction analysis, such as BRD4, TOP2A, and TRIM28. Bromodomain protein 4 (BRD4) is a histone acetylation reader and writer that plays an important role in DNA replication, transcription, and repair [73]. This epigene is critical for the maintenance of the higher-order chromatin structure, since its inhibition leads to chromatin decondensation and fragmentation, and it also can stimulate innate antiviral immunity [73]. BRD4 complexes with RelA and CDK9 and is functionally required for effective activation of NF-kB-dependent immediate early cytokine genes in response to viral patterns. In this sense, our results show a protein interaction with EP300, which involves the p300 / CBP complex, which is one of the best characterized cofactors of NF-kB and binds specifically to RelA [74]. Validating the possible importance of this system in infection with SARS-CoV2. Examples like this suggest that the virus, through these epigenetic remodelers, is promoting that the cell induces a chromatin remodeling that in many cases will lead to opening, both at the local level of specific genes and at the global level. Accordingly, an indicator of global changes is the increased expression of transcripts from repeated sequences such as LINE1. If this is so, then the virus is manipulating the chromatin aperture to promote the expression of genes that support its invasion. In this regard, other work has suggested the importance of LINE1 elements. Where this type of repetitive elements are very relevant in gene regulation, especially when these elements are in proximity to neighboring genes, since they could alter their expression. Therefore, the dysregulation of repeated elements such as LINE1 could indirectly change the cellular transcriptome [75].

Furthermore, we find epigenes that interact with the viral proteins directly or very closely. This connection suggests a virus-promoted modulation to affect the epigenome of the host cell's interactome. Which reinforces the idea that the virus strategy is partly to take advantage of the epigenetic machinery. In general, our data suggest that the SARS-CoV-2 infection deregulates the epigenetic master machinery. One of the points that should be taken into consideration in the future is that if this epigenetic machinery is not re-established after disease courses it could generate other diseases such as cancer in the long term. This is based on the fact that many of the genes that we found in our study have been proposed as epigenetic hallmarks in various neoplasms.

Our last key finding is the identification of driver epigenetic proteins and TFs involved in SARS-CoV-2 infection that can be targeted by existing drugs. We identified S-adenosyl-L-homocysteine (SAH) targeting several epigenetic components of the host response to SARS-CoV-2 infection. SAH is the product of the chemical reaction performed by methyltransferases using nucleic acids or proteins as substrates, and has been previously suggested as a potential treatment for viral infections such as ZIKA, MERS-CoV and SARS-CoV [76–79] due to its inhibitory activity of the viral RNA cap 2′-O-methyltransferase, formed by the NSP16-NSP10 complex [80,81]. Furthermore, given the interaction between DNMT1 and ORF8 at the protein level, SAH could potentially work against SARS-CoV-2 infection, not only by inhibiting the methyltransferase activity of NSP16-NSP10, but also by directly modulating the activity of the key host proteins involved in the transcriptional response to infection or by interfering with the interactions observed between ORF8 and DNMT1.

Furthermore, as anticipated, many proteins with epigenetic functions involved in SARS-CoV-2 infection have kinase activity and can be targeted by kinase inhibitors. One important example is imatinib, which we identified as a potential drug for SARS-CoV-2 and SARS-CoV, and is currently undergoing clinical trials to evaluate its efficacy in COVID-19 patients (NCT04394416, NCT04422678, NCT04346147 and NCT04357613; www.clinicaltrials.gov). Similarly, we found quercetin targeting several epifactors with kinase activity. Quercetin is a plant-derived compound with anti-inflammatory and antiviral effects [82,83] that has been evaluated in clinical trials as a dietary supplement or prophylaxis for COVID-19 (NCT04578158, NCT04377789 and NCT0446813). Even though some independent studies show no clear evidence of its effectiveness, preliminary data shows that it could be effective to decrease the frequency and duration of respiratory tract infections [84–86]. It is worth mentioning that these drugs are being tested in clinical trials based on their described inhibitory activity of enzymes related to the activation of immune response and inflammation, such as growth receptors [87]. The latter, together with our results, suggests that drugs targeting epigenetic mechanisms could be also effective to treat SARS-CoV-2 by modulating their kinase activity.

Finally, we also identified Bortezomib and betulinic acid associated with RELA. Bortezomib is a proteasome inhibitor that has been proposed as COVID-19 therapy given its capacity to inhibit (although only marginally) the papain-like protease (NSP3) of SARS-CoV, which also has deubiquitinase activity [88–90]. Likewise, betulinic acid has been proposed as a target of NSP3 in SARS-CoV-2 [91].

Together, we have supporting evidence that current drug-based therapies to treat COVID-19 also target the transcriptional response to infection by the modulation of the epigenetic proteins identified in this study. Furthermore, we provide additional new potential drug targets and drug candidates which could be effective and whose potential use has not been exploited yet (Figure 5). These results provide comprehensive evidence that epigenetic therapy could aid in restoring the transcriptional changes observed during infection. By using epigenetic drugs, a therapeutic effect can be achieved due to their systemic effects, which can be advantageous to treat a disease that targets different tissues and cellular mechanisms, as observed in COVID-19.

In this study, we used a blend of bioinformatic approaches to comparatively analyze transcriptomic data from SARS-CoV-2, SARS-CoV and MERS-CoV pulmonar infected cell lines and COVID-19 patient-derived samples.In particular, we focused on the epigenetic processes and transcriptional factors, since these have been widely proposed as the master regulators of the expression of most genes. We found that the transcriptional response elicited by MERS-CoV and SARS-CoV-2 is more similar to that observed for SARS-CoV, which is observed both in infected cell lines and in samples from patients with COVID-19. At the same time, we identified specific altered modules in the response to infection with SARS-CoV2 that could serve as an aliasing for the proposal of different therapeutic strategies based on epigenetic therapy. Thus, our results add a piece to the puzzle of the strategies used by the different coronaviruses to manipulate the gene regulation capacity of the cell. Although the pathways are differential between them, the virus objective is to take advantage of the TFs and various chromatin remodelers to avoid being detected and prevail in the invasion. This is a very fine strategy that the virus uses and it has been poorly studied in both its biological importance and its future therapeutic application. This could open a new window of opportunities for treatment and thus close the chapter on this pandemic disease.

## ACKNOWLEDGEMENTS

This work was supported by Apoyo para proyectos de investigación científica, desarrollo tecnológico e innovación en salud ante la contingencia por COVID-19, CONACyT [00312021 to ESR], Fondo CB-SEP-CONACyT [284748 to E.S.R.] and DSA-SEP [47310681 to E.S.R.]. M.S.A. and A.D.M.M. are doctoral students in the "Programa de Doctorado en Ciencias Bioquímicas, UNAM" and received a fellowship funding from CONACYT (M.S.A. CVU659273 and A.D.M.M. CVU894530). M.S.A. was awarded by the German Academic Exchange Service, DAAD (ref. 91693321). N.A. would like to acknowledge the Independent Research Fund Denmark (6108-00038B). E.S.R. was supported by the Departamento de Ciencias Naturales, UAM-Cuajimalpa.

**AUTHOR CONTRIBUTIONS**

M.S.A, E.I.N.D. and A.D.M.M. equally contributed to the data collection, bioinformatic analyses and manuscript writing. N.A. and J.B. provided critical feedback and helped to improve the manuscript. R.G.B. and E.S.R. were in charge of overall direction, planning, and supervision.

**COMPETING INTERESTS**

The authors declare no competing interests.

**REFERENCES**

1. Payne, S. Chapter 17 - Family Coronaviridae. in *Viruses* (ed. Payne, S.) 149–158 (Academic Press, 2017).

2. Memish, Z. A., Perlman, S., Van Kerkhove, M. D. & Zumla, A. Middle East respiratory syndrome. *Lancet* **395**, 1063–1077 (2020).

3. Hui, D. S. C. & Zumla, A. Severe Acute Respiratory Syndrome: Historical, Epidemiologic, and Clinical Features. *Infect. Dis. Clin. North Am.* **33**, 869–889 (2019).

4. World Health Organization. https://www.who.int/emergencies/diseases/novel-coronavirus-2019.

5. Youssef, N., Budd, A. & Bielawski, J. P. Introduction to Genome Biology and Diversity. *Methods Mol. Biol.* **1910**, 3–31 (2019).

6. Marazzi, I. *et al.* Suppression of the antiviral response by an influenza histone mimic. *Nature* **483**, 428–433 (2012).

7. Flanagan, J. M. Host epigenetic modifications by oncogenic viruses. *Br. J. Cancer* **96**, 183–188 (2007).

8. Menachery, V. D. *et al.* MERS-CoV and H5N1 influenza virus antagonize antigen presentation by altering the epigenetic landscape. *Proc. Natl. Acad. Sci. U. S. A.* **115**, E1012–E1021 (2018).

9. Marcos-Villar, L., Pazo, A. & Nieto, A. Influenza Virus and Chromatin: Role of the CHD1 Chromatin Remodeler in the Virus Life Cycle. *J. Virol.* **90**, 3694–3707 (2016).

10. Marcos-Villar, L. *et al.* Epigenetic control of influenza virus: role of H3K79 methylation in interferon-induced antiviral response. *Sci. Rep.* **8**, 1230 (2018).

11. O'Connor, C. M., DiMaggio, P. A., Jr, Shenk, T. & Garcia, B. A. Quantitative proteomic discovery of dynamic epigenome changes that control human cytomegalovirus (HCMV) infection. *Mol. Cell. Proteomics* **13**, 2399–2410 (2014).

12. Menachery, V. D. *et al.* Pathogenic influenza viruses and coronaviruses utilize similar and contrasting approaches to control interferon-stimulated gene responses. *MBio* **5**, e01174–14 (2014).

13. Ochsner, S. A., Pillich, R. T. & McKenna, N. J. Consensus transcriptional regulatory networks of coronavirus-infected human cells. *Sci Data* **7**, 314 (2020).

14. Singh, K. *et al.* Network Analysis and Transcriptome Profiling Identify Autophagic and Mitochondrial Dysfunctions in SARS-CoV-2 Infection. *bioRxiv* (2020) doi:10.1101/2020.05.13.092536.

15. Ganesan, A., Arimondo, P. B., Rots, M. G., Jeronimo, C. & Berdasco, M. The timeline of epigenetic drug discovery: from reality to dreams. *Clin. Epigenetics* **11**, 174 (2019).

16. WHO Solidarity Trial Consortium *et al.* Repurposed antiviral drugs for COVID-19; interim WHO SOLIDARITY trial results. *Infectious Diseases (except HIV/AIDS)* (2020) doi:10.1101/2020.10.15.20209817.

17. Bolger, A. M., Lohse, M. & Usadel, B. Trimmomatic: a flexible trimmer for Illumina sequence data. *Bioinformatics* **30**, 2114–2120 (2014).

18. Andrews, S. & Others. FastQC: a quality control tool for high throughput sequence data. (2010).

19. Dobin, A. *et al.* STAR: ultrafast universal RNA-seq aligner. *Bioinformatics* **29**, 15–21 (2013).

20. Jin, Y., Tam, O. H., Paniagua, E. & Hammell, M. TEtranscripts: a package for including transposable elements in differential expression analysis of RNA-seq datasets. *Bioinformatics* **31**, 3593–3599 (2015).

21. Love, M. I., Huber, W. & Anders, S. Moderated estimation of fold change and dispersion for RNA-seq data with DESeq2. *Genome Biol.* **15**, 550 (2014).

22. Patro, R., Duggal, G., Love, M. I., Irizarry, R. A. & Kingsford, C. Salmon provides fast and bias-aware quantification of transcript expression. *Nat. Methods* **14**, 417–419 (2017).

23. Medvedeva, Y. A. *et al.* EpiFactors: a comprehensive database of human epigenetic factors and

complexes. *Database* **2015**, bav067 (2015).

24. Khare, S. P. *et al.* HIstome—a relational knowledgebase of human histone proteins and histone modifying enzymes. *Nucleic Acids Res.* **40**, D337–D342 (2011).

25. Singh Nanda, J., Kumar, R. & Raghava, G. P. S. dbEM: A database of epigenetic modifiers curated from cancerous and normal genomes. *Sci. Rep.* **6**, 19340 (2016).

26. Lambert, S. A. *et al.* The Human Transcription Factors. *Cell* **172**, 650–665 (2018).

27. Robinson, M. D., McCarthy, D. J. & Smyth, G. K. edgeR: a Bioconductor package for differential expression analysis of digital gene expression data. *Bioinformatics* **26**, 139–140 (2010).

28. Langfelder, P. & Horvath, S. WGCNA: an R package for weighted correlation network analysis. *BMC Bioinformatics* **9**, 559 (2008).

29. Yu, G., Wang, L.-G., Han, Y. & He, Q.-Y. clusterProfiler: an R Package for Comparing Biological Themes Among Gene Clusters. *OMICS* **16**, 284–287 (2012).

30. Raudvere, U. *et al.* g:Profiler: a web server for functional enrichment analysis and conversions of gene lists (2019 update). *Nucleic Acids Res.* **47**, W191–W198 (2019).

31. Gordon, D. E. *et al.* A SARS-CoV-2 protein interaction map reveals targets for drug repurposing. *Nature* **583**, 459–468 (2020).

32. Stukalov, A., Girault, V., Grass, V., Bergant, V. & Karayel, O. Multi-level proteomics reveals host-perturbation strategies of SARS-CoV-2 and SARS-CoV. *Biorxiv* (2020).

33. Kotlyar, M., Pastrello, C., Malik, Z. & Jurisica, I. IID 2018 update: context-specific physical protein-protein interactions in human, model organisms and domesticated species. *Nucleic Acids Res.* **47**, D581–D589 (2019).

34. Horvath, S. & Langfelder, P. Tutorials for the WGCNA package for R: WGCNA Background and glossary. (2011).

35. Csárdi, G. & Nepusz, T. The igraph software package for complex network research. *InterJournal, complex systems* **1695**, 1–9 (2006).

36. Alcaraz, N. *et al.* Robust de novo pathway enrichment with KeyPathwayMiner 5. *F1000Res.* **5**, 1531 (2016).

37. Sadegh, S. *et al.* Exploring the SARS-CoV-2 virus-host-drug interactome for drug repurposing. *Nat.*

*Commun.* **11**, 3518 (2020).

38. Rayner, R. E., Makena, P., Prasad, G. L. & Cormet-Boyaka, E. Optimization of Normal Human Bronchial Epithelial (NHBE) Cell 3D Cultures for in vitro Lung Model Studies. *Sci. Rep.* **9**, 500 (2019).

39. Davis, A. S. *et al.* Validation of normal human bronchial epithelial cells as a model for influenza A infections in human distal trachea. *J. Histochem. Cytochem.* **63**, 312–328 (2015).

40. Bulut-Karslioglu, A. *et al.* Suv39h-dependent H3K9me3 marks intact retrotransposons and silences LINE elements in mouse embryonic stem cells. *Mol. Cell* **55**, 277–290 (2014).

41. Castro-Diaz, N. *et al.* Evolutionally dynamic L1 regulation in embryonic stem cells. *Genes Dev.* **28**, 1397–1409 (2014).

42. Blanco-Melo, D. *et al.* Imbalanced Host Response to SARS-CoV-2 Drives Development of COVID-19. *Cell* **181**, 1036–1045.e9 (2020).

43. Schultz, D. C., Ayyanathan, K., Negorev, D., Maul, G. G. & Rauscher, F. J., 3rd. SETDB1: a novel KAP-1-associated histone H3, lysine 9-specific methyltransferase that contributes to HP1-mediated silencing of euchromatic genes by KRAB zinc-finger proteins. *Genes Dev.* **16**, 919–932 (2002).

44. Hess, J., Angel, P. & Schorpp-Kistner, M. AP-1 subunits: quarrel and harmony among siblings. *J. Cell Sci.* **117**, 5965–5973 (2004).

45. Fagone, P. *et al.* Transcriptional landscape of SARS-CoV-2 infection dismantles pathogenic pathways activated by the virus, proposes unique sex-specific differences and predicts tailored therapeutic strategies. *Autoimmun. Rev.* **19**, 102571 (2020).

46. Bhatt, D. & Ghosh, S. Regulation of the NF-κB-Mediated Transcription of Inflammatory Genes. *Front. Immunol.* **5**, 71 (2014).

47. Hu, Y. *et al.* The Severe Acute Respiratory Syndrome Coronavirus Nucleocapsid Inhibits Type I Interferon Production by Interfering with TRIM25-Mediated RIG-I Ubiquitination. *J. Virol.* **91**, (2017).

48. Fehervari, Z. Putting a KAP on infection. *Nat. Immunol.* **12**, 816–816 (2011).

49. Pavlova, N. I., Savinova, O. V., Nikolaeva, S. N., Boreko, E. I. & Flekhter, O. B. Antiviral activity of betulin, betulinic and betulonic acids against some enveloped and non-enveloped viruses. *Fitoterapia* **74**, 489–492 (2003).

50. Aiken, C. & Chen, C. H. Betulinic acid derivatives as HIV-1 antivirals. *Trends Mol. Med.* **11**, 31–36

(2005).

51. Gupta, S. C., Patchva, S. & Aggarwal, B. B. Therapeutic roles of curcumin: lessons learned from clinical trials. *AAPS J.* **15**, 195–218 (2013).

52. Bollati, V. & Baccarelli, A. Environmental epigenetics. *Heredity* **105**, 105–112 (2010).

53. Beacon, T. H., Su, R.-C., Lakowski, T. M., Delcuve, G. P. & Davie, J. R. SARS-CoV-2 multifaceted interaction with the human host. Part II: Innate immunity response, immunopathology, and epigenetics. *IUBMB Life* (2020) doi:10.1002/iub.2379.

54. Chu, H. *et al.* Comparative Replication and Immune Activation Profiles of SARS-CoV-2 and SARS-CoV in Human Lungs: An Ex Vivo Study With Implications for the Pathogenesis of COVID-19. *Clinical Infectious Diseases* vol. 71 1400–1409 (2020).

55. Liu, Y. *et al.* TRIM25 Promotes TNF-α–Induced NF-κB Activation through Potentiating the K63-Linked Ubiquitination of TRAF2. *The Journal of Immunology* **204**, 1499–1507 (2020).

56. Wolter, S. *et al.* c-Jun Controls Histone Modifications, NF-κB Recruitment, and RNA Polymerase II Function To Activate the ccl2 Gene. *Mol. Cell. Biol.* **28**, 4407–4423 (2008).

57. Ndlovu, 'matladi N. *et al.* Hyperactivated NF-{kappa}B and AP-1 transcription factors promote highly accessible chromatin and constitutive transcription across the interleukin-6 gene promoter in metastatic breast cancer cells. *Mol. Cell. Biol.* **29**, 5488–5504 (2009).

58. Bektas, A. *et al.* Age-associated changes in basal NF-κB function in human CD4+ T lymphocytes via dysregulation of PI3 kinase. *Aging* **6**, 957–974 (2014).

59. Do, L. A. H., Anderson, J., Mulholland, E. K. & Licciardi, P. V. Can data from paediatric cohorts solve the COVID-19 puzzle? *PLoS Pathog.* **16**, e1008798 (2020).

60. Martinez-Moczygemba, M., Gutch, M. J., French, D. L. & Reich, N. C. Distinct STAT Structure Promotes Interaction of STAT2 with the p48 Subunit of the Interferon-α-stimulated Transcription Factor ISGF3. *J. Biol. Chem.* **272**, 20070–20076 (1997).

61. Robichon, K. *et al.* Identification of Interleukin1β as an Amplifier of Interferon alpha-induced Antiviral Responses. *PLoS Pathog.* **16**, e1008461 (2020).

62. Hadjadj, J. *et al.* Impaired type I interferon activity and inflammatory responses in severe COVID-19 patients. *Science* **369**, 718–724 (2020).

63. Prasad, K. *et al.* Targeting hub genes and pathways of innate immune response in COVID-19: A network biology perspective. *Int. J. Biol. Macromol.* **163**, 1–8 (2020).

64. Zhang, Q. *et al.* Inborn errors of type I IFN immunity in patients with life-threatening COVID-19. *Science* (2020) doi:10.1126/science.abd4570.

65. Ciancanelli, M. J. *et al.* Infectious disease. Life-threatening influenza and impaired interferon amplification in human IRF7 deficiency. *Science* **348**, 448–453 (2015).

66. Pinto, B. G. G. *et al.* ACE2 Expression Is Increased in the Lungs of Patients With Comorbidities Associated With Severe COVID-19. *J. Infect. Dis.* **222**, 556–563 (2020).

67. Clarke, N. E., Belyaev, N. D., Lambert, D. W. & Turner, A. J. Epigenetic regulation of angiotensin-converting enzyme 2 (ACE2) by SIRT1 under conditions of cell energy stress. *Clin. Sci.* **126**, 507–516 (2014).

68. Dell'Omo, G. *et al.* Inhibition of SIRT1 deacetylase and p53 activation uncouples the anti-inflammatory and chemopreventive actions of NSAIDs. *Br. J. Cancer* **120**, 537–546 (2019).

69. Sawalha, A. H., Zhao, M., Coit, P. & Lu, Q. Epigenetic dysregulation of ACE2 and interferon-regulated genes might suggest increased COVID-19 susceptibility and severity in lupus patients. *Clin. Immunol.* **215**, 108410 (2020).

70. Wang, X. *et al.* Hepatitis B virus X protein induces hepatic stem cell-like features in hepatocellular carcinoma by activating KDM5B. *World J. Gastroenterol.* **23**, 3252–3261 (2017).

71. Wu, L. *et al.* KDM5 histone demethylases repress immune response via suppression of STING. *PLoS Biol.* **16**, e2006134 (2018).

72. Zhang, X., Liu, L., Yuan, X., Wei, Y. & Wei, X. JMJD3 in the regulation of human diseases. *Protein Cell* **10**, 864–882 (2019).

73. Wang, J. *et al.* BRD4 inhibition exerts anti-viral activity through DNA damage-dependent innate immune responses. *PLoS Pathog.* **16**, e1008429 (2020).

74. Tian, B. *et al.* BRD4 Couples NF-κB/RelA with Airway Inflammation and the IRF-RIG-I Amplification Loop in Respiratory Syncytial Virus Infection. *J. Virol.* **91**, (2017).

75. Faulkner, G. J. *et al.* The regulated retrotransposon transcriptome of mammalian cells. *Nat. Genet.* **41**, 563–571 (2009).

76. Bray, M., Driscoll, J. & Huggins, J. W. Treatment of lethal Ebola virus infection in mice with a single dose of an S-adenosyl-L-homocysteine hydrolase inhibitor. *Antiviral Res.* **45**, 135–147 (2000).

77. Coutard, B. *et al.* Zika Virus Methyltransferase: Structure and Functions for Drug Design Perspectives. *J. Virol.* **91**, (2017).

78. Aouadi, W. *et al.* Binding of the Methyl Donor S-Adenosyl-l-Methionine to Middle East Respiratory Syndrome Coronavirus 2'-O-Methyltransferase nsp16 Promotes Recruitment of the Allosteric Activator nsp10. *J. Virol.* **91**, (2017).

79. Li, G. & De Clercq, E. Therapeutic options for the 2019 novel coronavirus (2019-nCoV). *Nat. Rev. Drug Discov.* **19**, 149–150 (2020).

80. Mahalapbutr, P., Kongtaworn, N. & Rungrotmongkol, T. Structural insight into the recognition of S-adenosyl-L-homocysteine and sinefungin in SARS-CoV-2 Nsp16/Nsp10 RNA cap 2'-O-Methyltransferase. *Comput. Struct. Biotechnol. J.* **18**, 2757–2765 (2020).

81. Krafcikova, P., Silhan, J., Nencka, R. & Boura, E. Structural analysis of the SARS-CoV-2 methyltransferase complex involved in RNA cap creation bound to sinefungin. *Nat. Commun.* **11**, 3717 (2020).

82. Uchide, N. & Toyoda, H. Antioxidant therapy as a potential approach to severe influenza-associated complications. *Molecules* **16**, 2032–2052 (2011).

83. Nair, M. P. N. *et al.* The flavonoid, quercetin, differentially regulates Th-1 (IFNγ) and Th-2 (IL4) cytokine gene expression by normal peripheral blood mononuclear cells. *Biochimica et Biophysica Acta (BBA) - Molecular Cell Research* **1593**, 29–36 (2002).

84. Karunakaran, K. B., Balakrishnan, N. & Ganapathiraju, M. Potentially repurposable drugs for COVID-19 identified from SARS-CoV-2 Host Protein Interactome. *Res Sq* (2020) doi:10.21203/rs.3.rs-30363/v1.

85. Colunga Biancatelli, R. M. L., Berrill, M., Catravas, J. D. & Marik, P. E. Quercetin and Vitamin C: An Experimental, Synergistic Therapy for the Prevention and Treatment of SARS-CoV-2 Related Disease (COVID-19). *Front. Immunol.* **11**, 1451 (2020).

86. Aucoin, M. *et al.* The effect of quercetin on the prevention or treatment of COVID-19 and other respiratory tract infections in humans: A rapid review. *Adv Integr Med* (2020)

doi:10.1016/j.aimed.2020.07.007.

87. Luo, W. *et al.* Targeting JAK-STAT Signaling to Control Cytokine Release Syndrome in COVID-19. *Trends Pharmacol. Sci.* **41**, 531–543 (2020).

88. Lindner, H. A. *et al.* The papain-like protease from the severe acute respiratory syndrome coronavirus is a deubiquitinating enzyme. *J. Virol.* **79**, 15199–15208 (2005).

89. Schneider, M. *et al.* Severe acute respiratory syndrome coronavirus replication is severely impaired by MG132 due to proteasome-independent inhibition of M-calpain. *J. Virol.* **86**, 10112–10122 (2012).

90. Longhitano, L. *et al.* Proteasome Inhibitors as a Possible Therapy for SARS-CoV-2. *Int. J. Mol. Sci.* **21**, (2020).

91. Chojnacka, K., Witek-Krowiak, A., Skrzypczak, D., Mikula, K. & Młynarz, P. Phytochemicals containing biologically active polyphenols as an effective agent against Covid-19-inducing coronavirus. *J. Funct. Foods* **73**, 104146 (2020).

**FIGURE LEGENDS**

Figure 1. Differential expression analysis of coronavirus-infected cell lines. **a** Intersection size of the DEGs common to each viral infection represented as single dots (virus-associated gene sets) and the size of their intersections with the other sets (multiple vertical dots). **b** Top 10 simplified enriched Gene Ontology terms of biological process in the virus-associated gene sets ordered by q-value. **c** Shared differentially expressed epigenes between virus-associated gene sets; text color corresponds to the gene classification as either TF (red) or epifactor (blue) (upper panel). Log2 fold change of shared differentially expressed epifactors in each cell line are also shown as a heatmap (lower panel); blank color represents non-significant differential expression, text highlight corresponds to the intersections shown in the Venn diagram. **d** Functional classification of the identified epifactos; text color corresponds to the intersection color of **c**. **e** Characterization of the DNA-binding domain (DBDs) of human transcription factors (TFs) altered by the viral infection of coronaviruses.

Figure 2. Differential expression analysis of COVID-19 patient samples. **a** Number of shared differentially expressed genes between different tissue samples. **b** Log2 fold change of shared differentially expressed genes in patients samples. **c** Top 10 simplified enriched Gene Ontology terms of biological process in the patient samples ordered by q-value. **d** Shared differentially expressed genes between SARS-CoV-2 infected cell lines and COVID-19 patient samples. Epifactors and transcription factors in common are denoted by text color.

Figure 3. Relevant modules for coronaviruses infection. Summary of the analyses used to identify relevant modules for each infection. From left to right, grids show the module-trait correlation, the enrichment of epigenes, the enrichment of DEGs found in cell lines, enrichment of DEGs found in patient samples and information of the module size.

Figure 4. Protein-protein interactions network containing SARS-CoV-2-DEGs, patient-DEGs or selected epigenes for modules 4, 6, 8, 10 , 11 and 12.

Figure 5. Relevant epigenes in SARS-CoV-2 infection with therapeutic potential. Epigenetic targets are indicated in different processes such as nucleosome occupancy (1), histone modification (2), DNA methylation (3) and also transcription factors (4). Top gene candidate targets are highlighted in red. Created with BioRender.com

Table 1. Drugs targeting candidate epigenes from modules 4, 6, 8, 10, 11 and 12.

| Target protein | Drug Name | Module | Function |
|---|---|---|---|
| JUN | Vinblastine, Pseudoephedrine, LGD-1550, 4-{[5-chloro-4-(1H-indol-3-yl)pyrimidin-2-yl]amino}-N-ethylpiperidine-1-carboxamide, Irbesartan, Arsenic trioxide | module 8 | TF |
| TOP2A | Genistein, Fluorouracil, Intoplicine, Enoxacin, Sparfloxacin, Amrubicin, Etoposide, Epirubicin, Ciprofloxacin, Myricetin, Mitoxantrone, Trovafloxacin, RTA 744, Daunorubicin, Norfloxacin, Finafloxacin, Dexrazoxane, 13-deoxydoxorubicin, Idarubicin, Lomefloxacin, Lucanthone, Pefloxacin, Valrubicin, Amsacrine, Levofloxacin, Doxorubicin, Declopramide, Annamycin, Banoxantrone, ZEN-012, Podofilox, Aldoxorubicin, Teniposide, Moxifloxacin, SP1049C, Amonafide, Dactinomycin, Fleroxacin, Becatecarin, Ofloxacin, Elsamitrucin | module 8 | Epifactor (Chromatin remodeling) |
| BRD4 | Fedratinib, Panobinostat, Romidepsin, Birabresib, Alprazolam, Vorinostat, Volasertib, Alobresib, Belinostat, Apabetalone | module 6 | Epifactor (Histone modification read) |
| EP300 | Curcumin | module 6 | Epifactor (Histone modification write) |
| DNMT1 | S-adenosyl-L-homocysteine, Procainamide, Palifosfamide, Cefalotin, Decitabine, Azacitidine, Flucytosine, Epigallocatechin gallate, Hydralazine | module 4 | Epifactor (DNA methylation) |
| SENP3 | Methylphenidate | module 4 | Epifactor (Histone modification erase, Histone modification write cofactor) |
| SIRT6 | 7-[4-(Dimethylamino)Phenyl]-N-Hydroxy-4,6-Dimethyl-7-Oxo-2,4-Heptadienamide | module 4 | Epifactor (Histone modification erase) |
| FOS | Pseudoephedrine, Nadroparin | module 12 | TF |
| RELA | SC-236, Betulinic Acid, Bortezomib, Dimethyl fumarate, PHENYL-5-(1H-PYRAZOL-3-YL)-1,3-THIAZOLE, Indoprofen | module 12 | TF |
| STAT1 | Epigallocatechin gallate | module 12 | TF |
| STAT5A | AZD-1480 | module 11 | TF |
| SMAD3 | Ellagic Acid | module 10 | TF |

Table 2. Candidate epigenes for drug development in modules 4, 6, 8, 10, 11 and 12.

| Target protein | Module | Function |
|---|---|---|
| MOV10 | module 12 | Epifactor (Chromatin remodeling) |
| MTA1 | module 4 | Epifactor (Chromatin remodeling cofactor) |
| TLE1 | module 12 | Epifactor (Chromatin remodeling, Histone modification cofactor) |
| TAF4 | module 12 | Epifactor (Histone chaperone) |
| BAP1 | module 4 | Epifactor (Histone modification erase, Polycomb group (PcG) protein) |
| TRIM28 | module 4 | Epifactor (Histone modification read) |
| PRDM1 | module 12 | Epifactor (Histone modification write cofactor) |
| SETDB1 | module 6 | Epifactor (Histone modification write) |
| UBE2DB8 | module 10 | Epifactor (Histone modification write) |
| PCGF5 | module 12 | Epifactor (Polycomb group (PcG) protein) |
| STAT2 | module 12 | TF |
| GMEB2 | module 6 | TF |
| ZNF277 | module 10 | TF |
| IRF7 | module 11 | TF |
| CEBPD | module 12 | TF |
| NR4A1 | module 12 | TF |
| ZNF652 | module 12 | TF |
| IRF2 | module 12 | TF |
| ZEB2 | module 12 | TF |
| SP110 | module 12 | TF |
| CUX1 | module 12 | TF |
| TRIM25 | module 12 | E3 ubiquitin ligase |

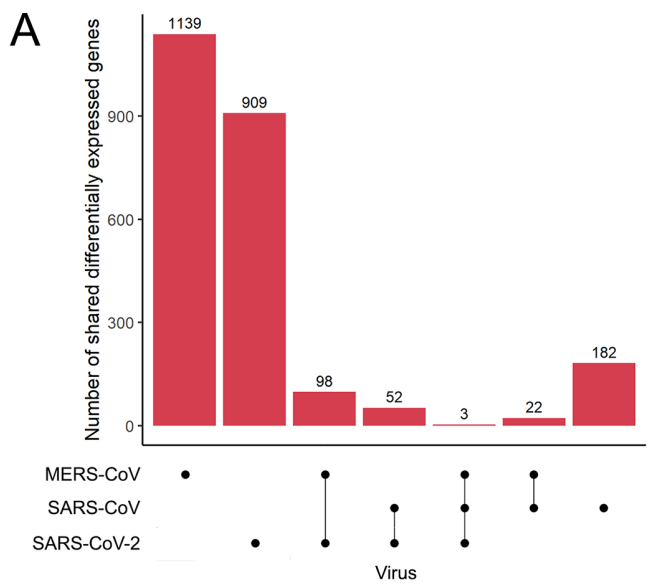
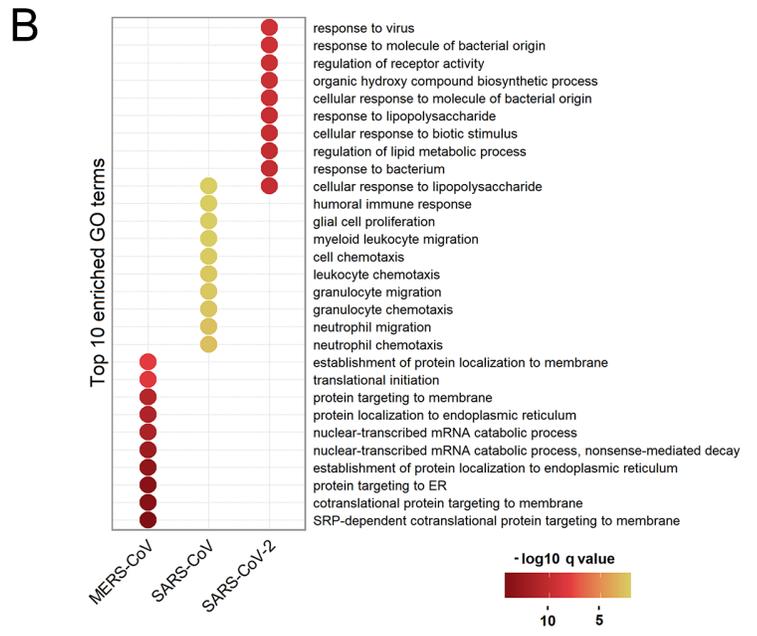
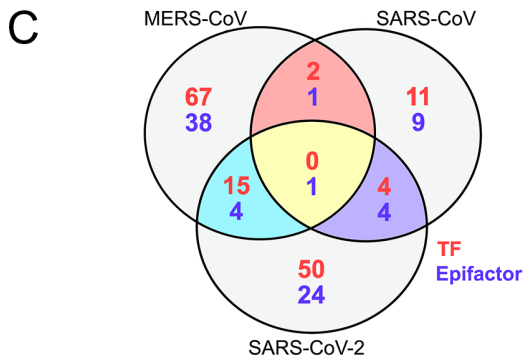
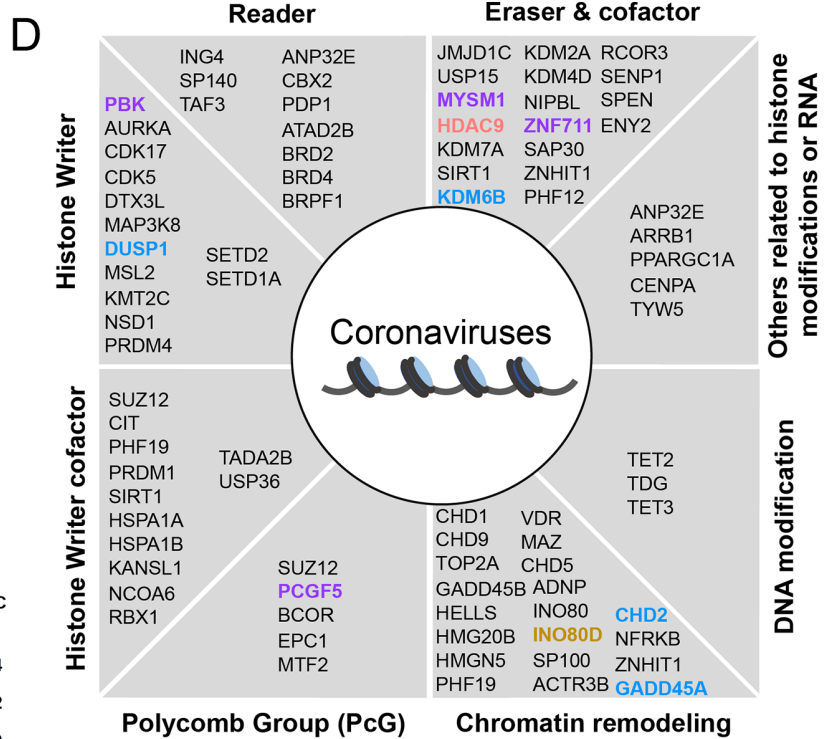
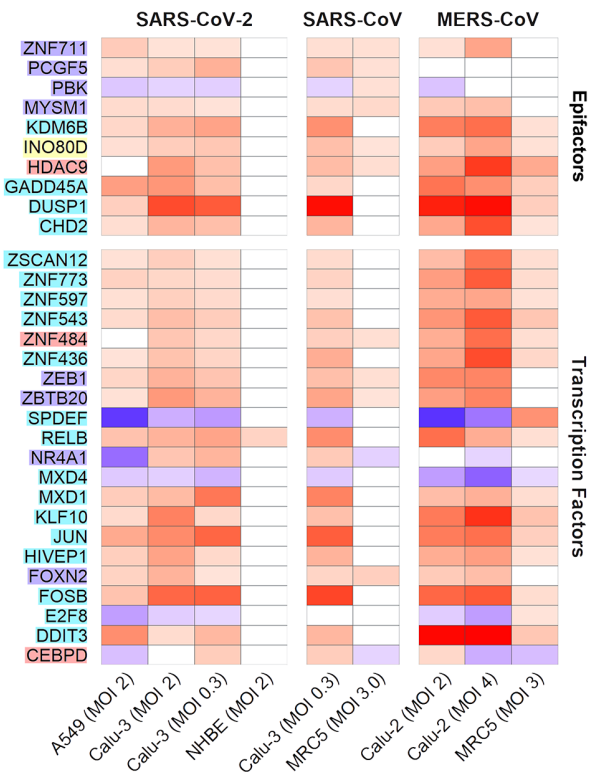
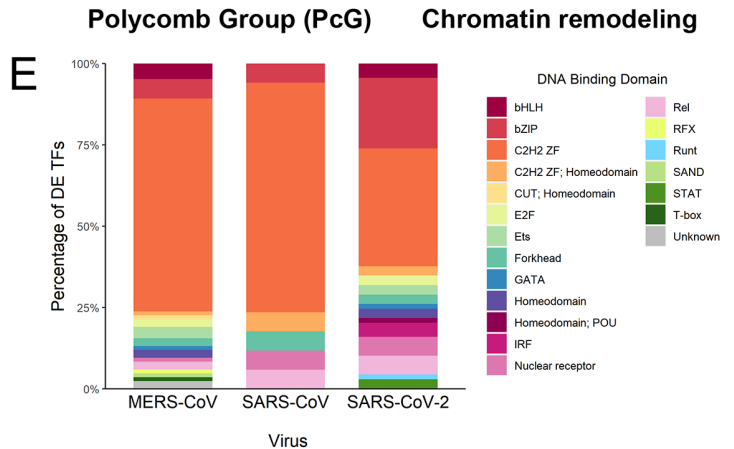

Figure 1

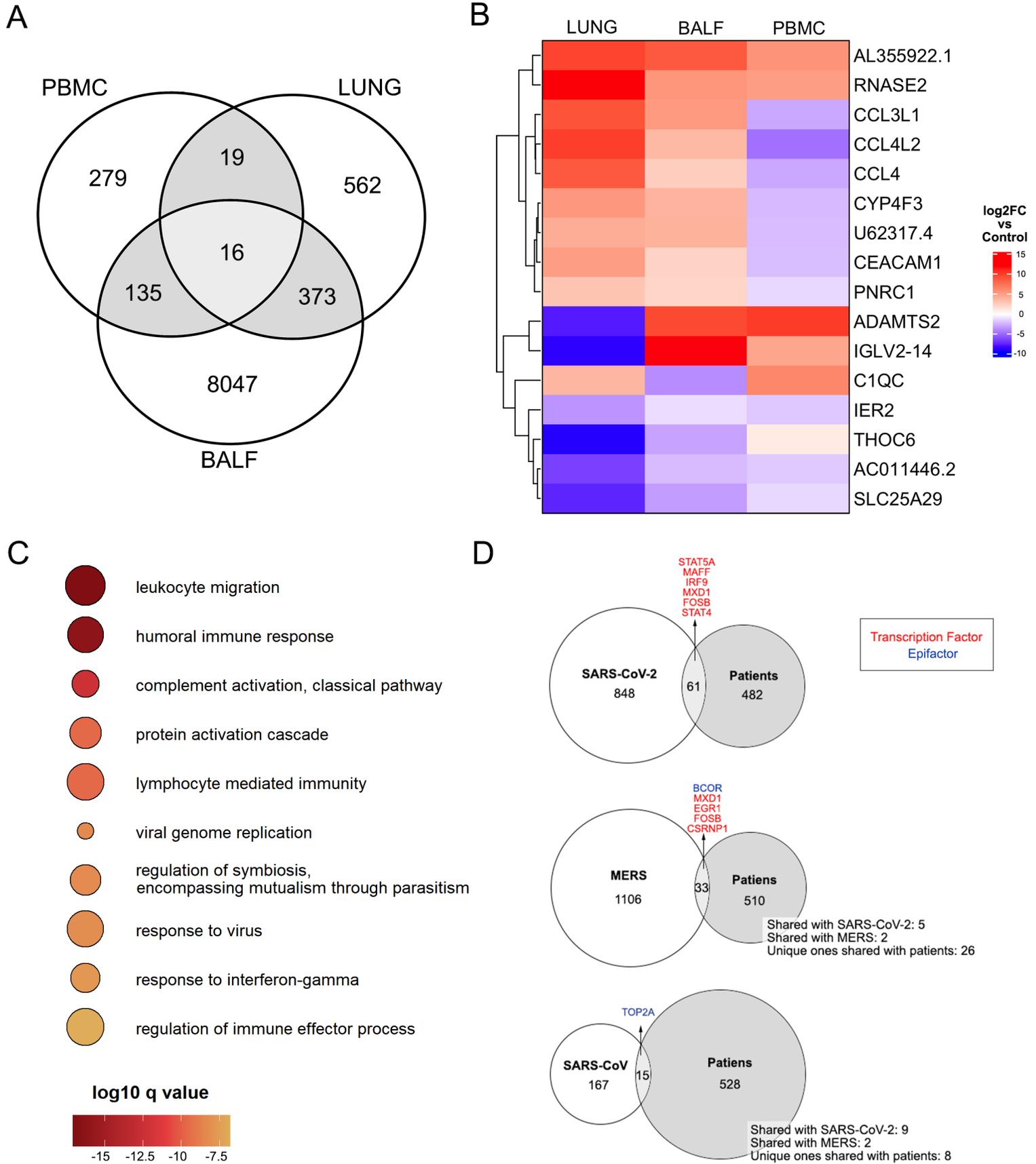

Figure 2

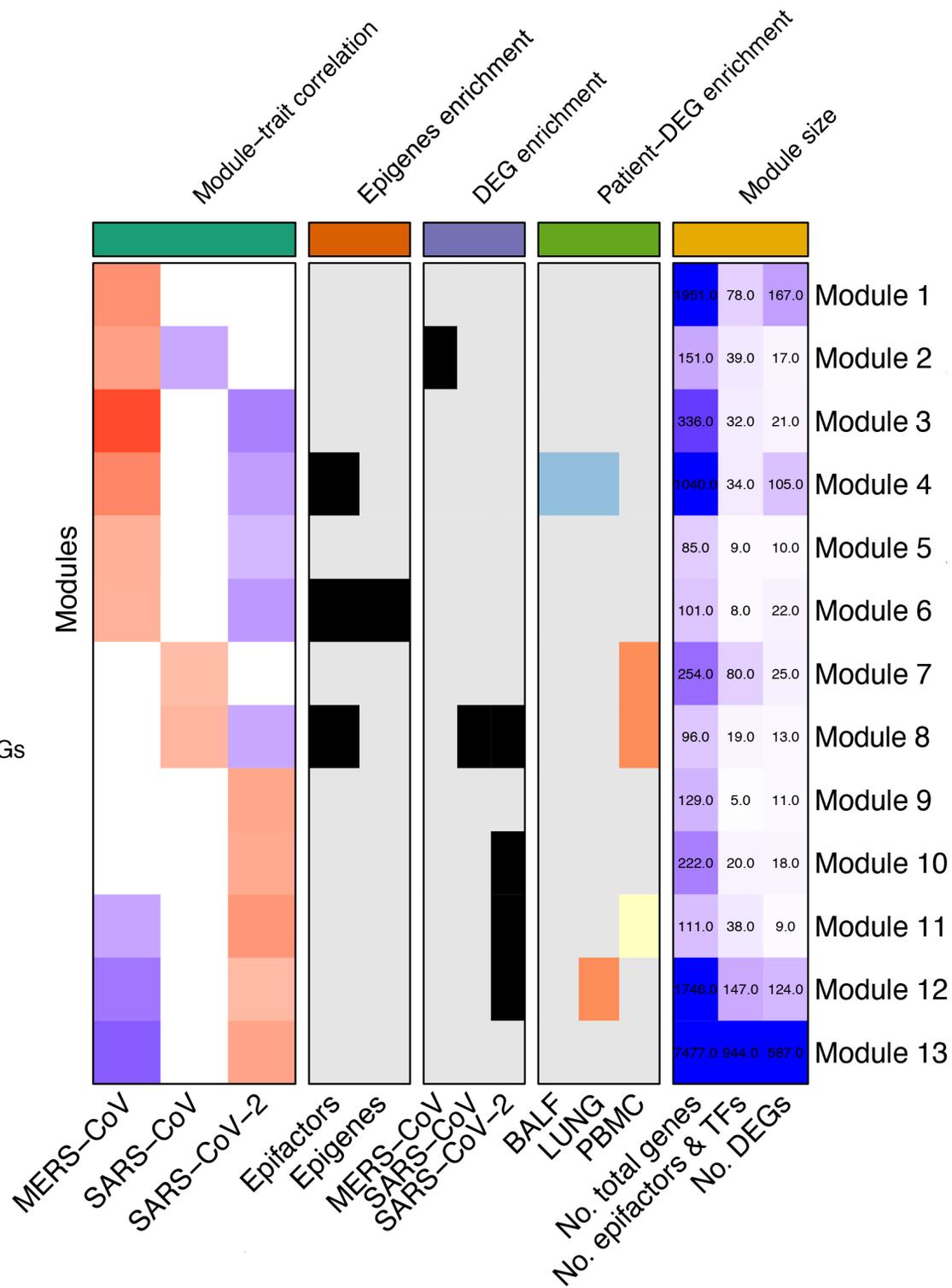

Figure 3

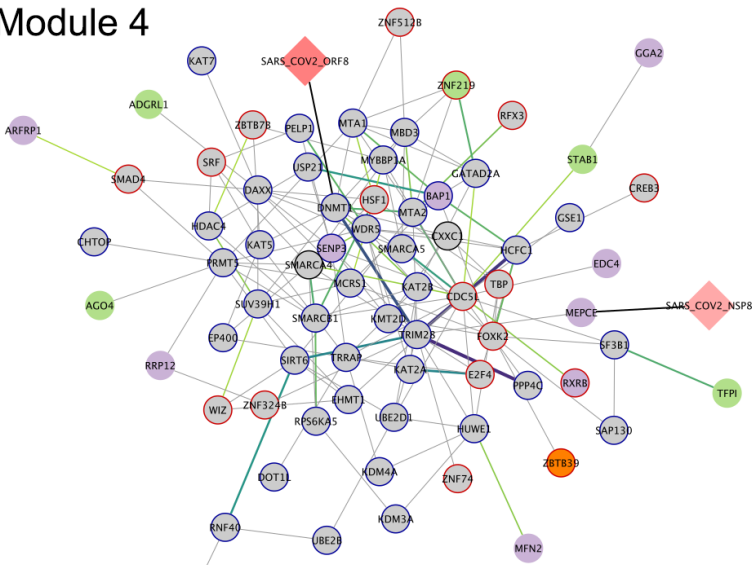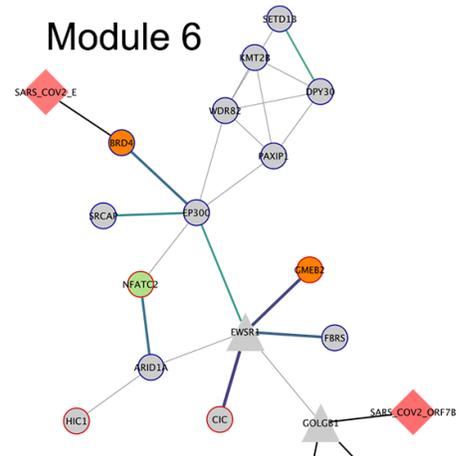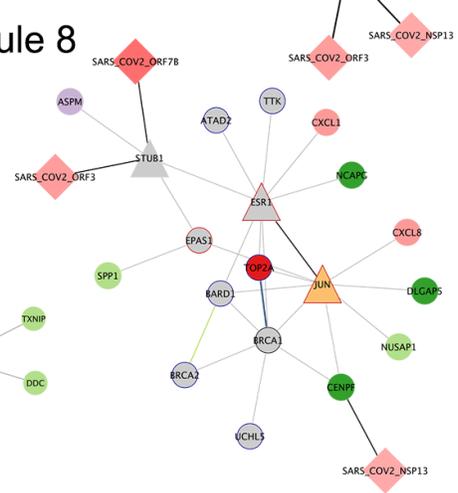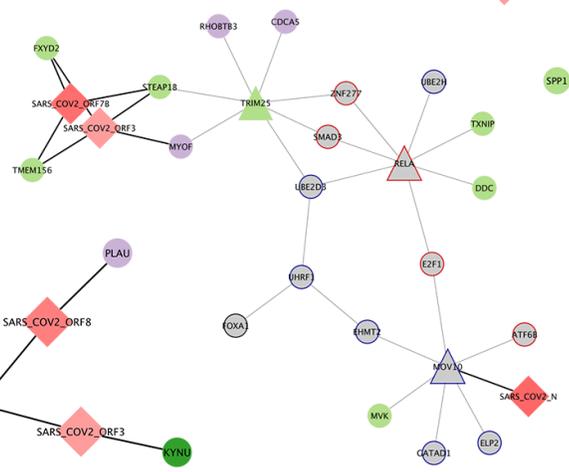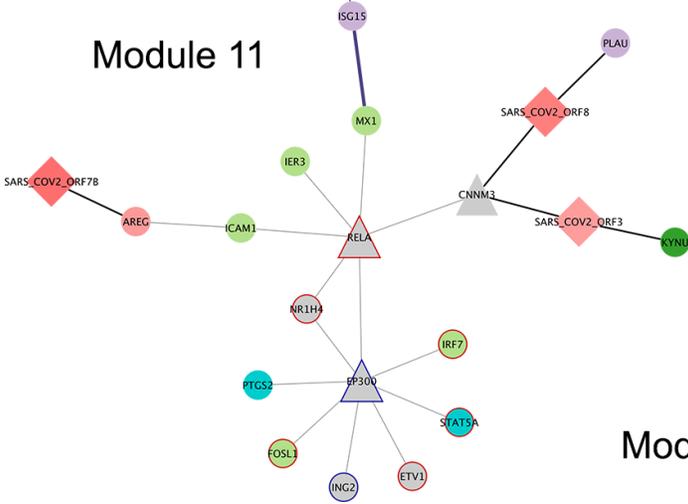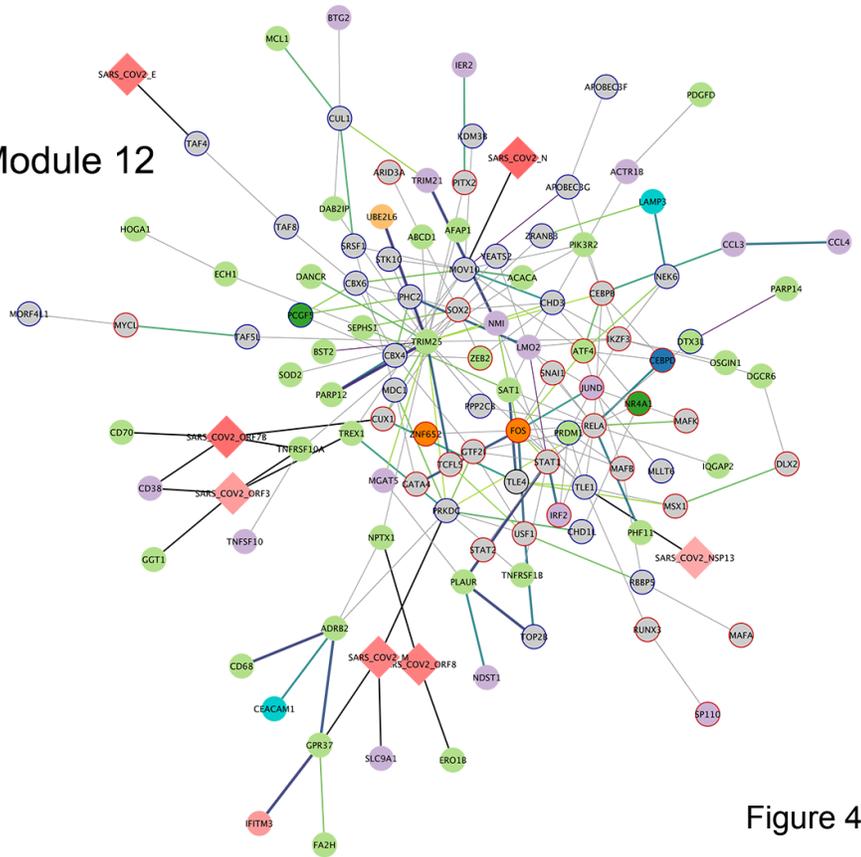

Figure 4

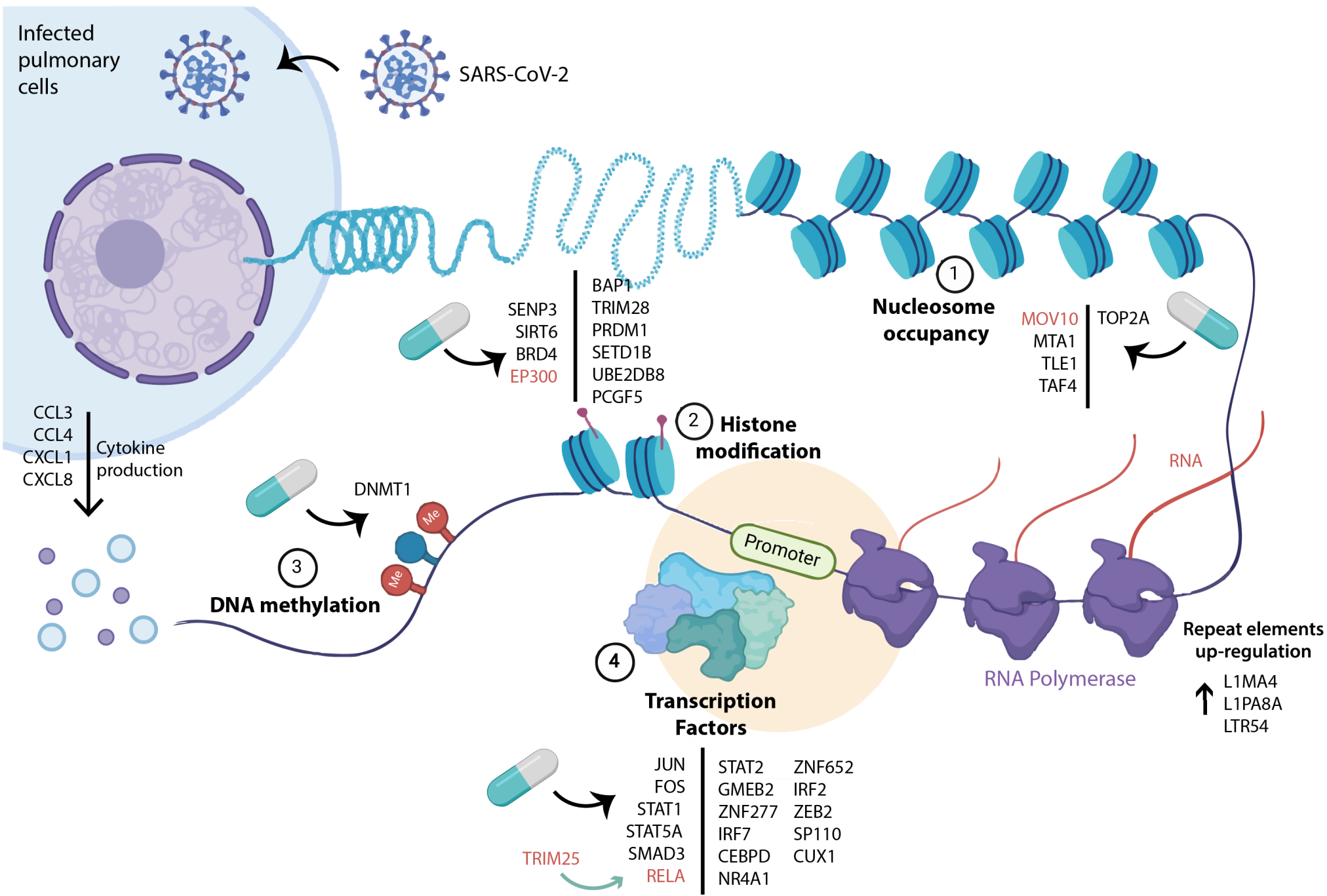

Figure 5

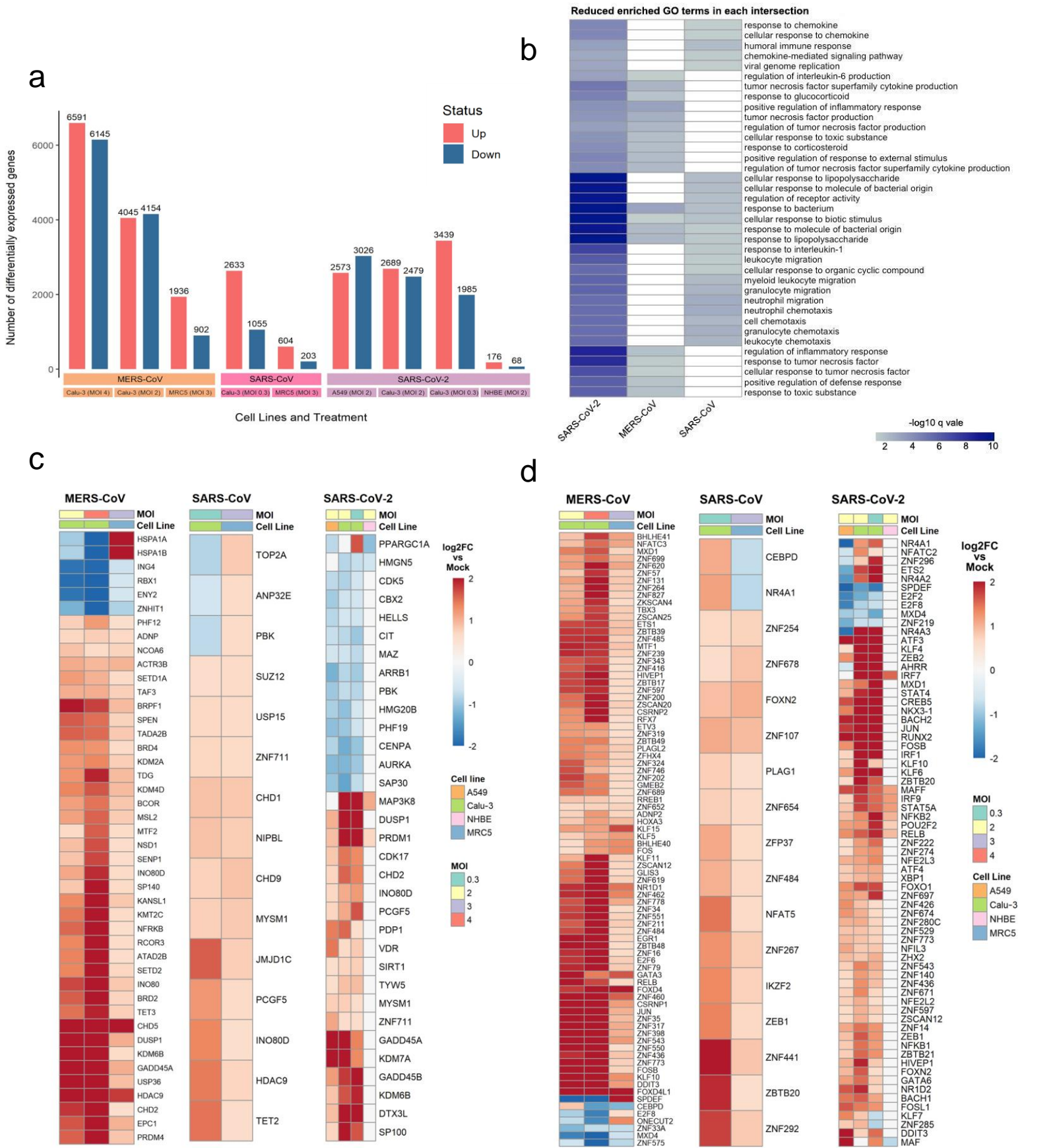

**Supplementary Figure 1. Differential expression analysis of Coronaviruses infected cell lines. a** Number of up and down-regulated differentially expressed genes (DEGs) in the analyzed cell lines for each viral infection. **b** Reduced enriched GO terms shared between two or more virus-associated gene sets. **c** Log2 fold-change expression of SARS-CoV-2 virus-associated epifactors across different viral infections; blank color represents non-significant differential expression. **d** Log2 fold-change expression of SARS-CoV-2 virus-associated transcription factors across different viral infections; blank color represents non-significant differential expression.

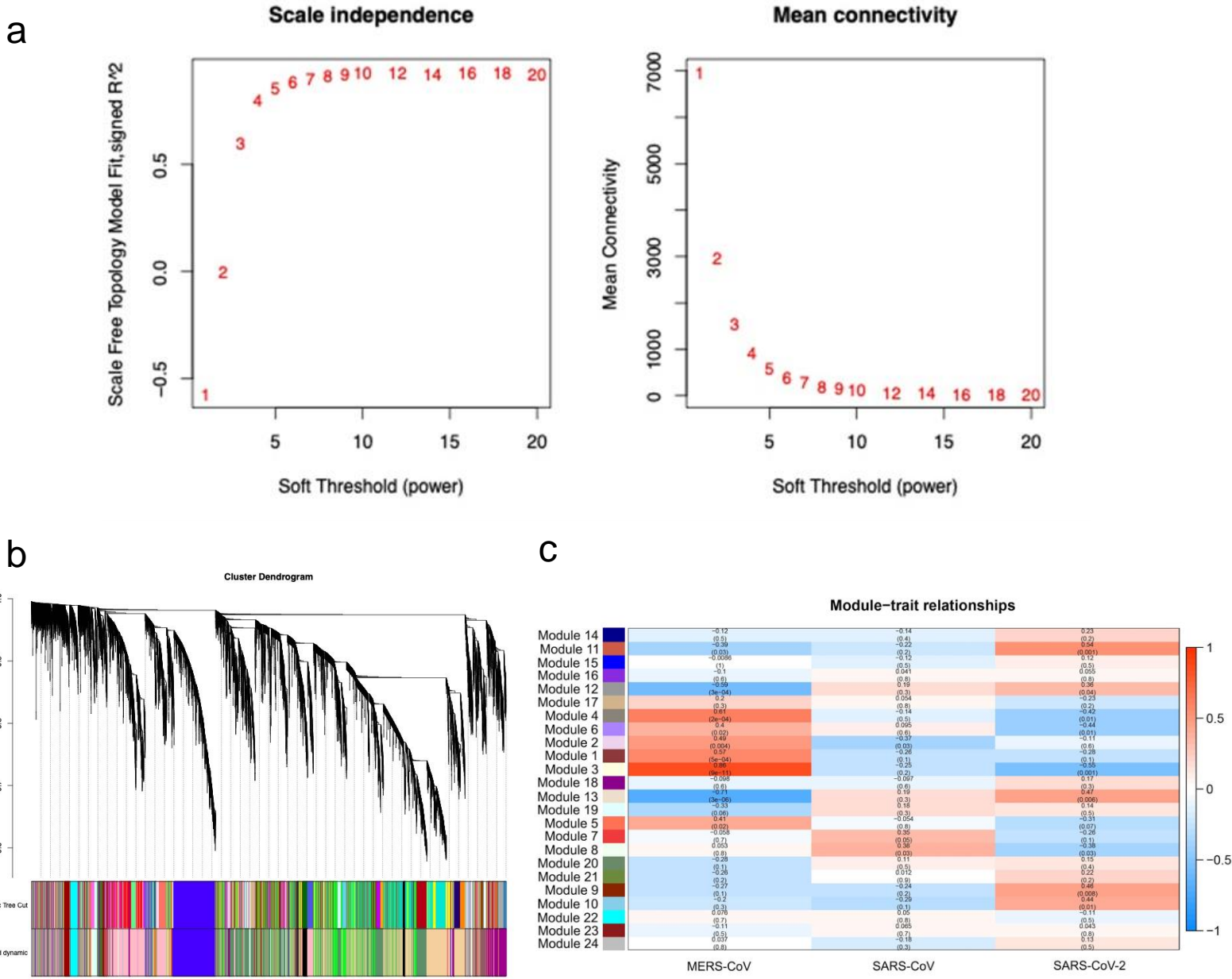

**Supplementary Figure 2. Weighted Gene Co-expression Network. a** Analysis of network topology for different soft-thresholding powers. **b** Clustering dendrogram of genes with their corresponding module colors before (Dynamic Cut Tree) and after (Merged Dynamic) similarity merging. **c** Module-virus association; the numbers inside each cell correspond to the correlation coefficient (top), which also dictates the cell color, and *p* value (bottom).

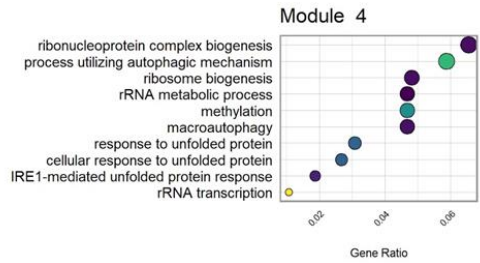
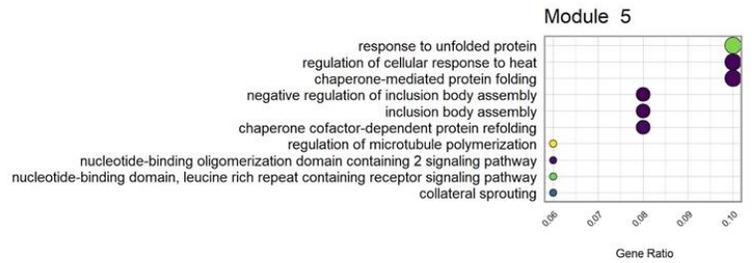
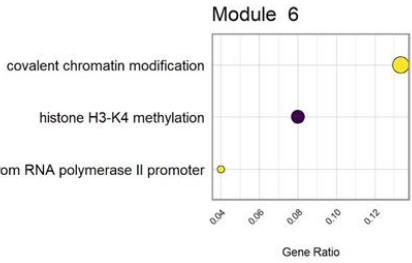
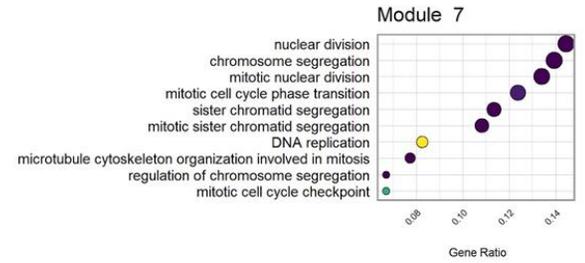
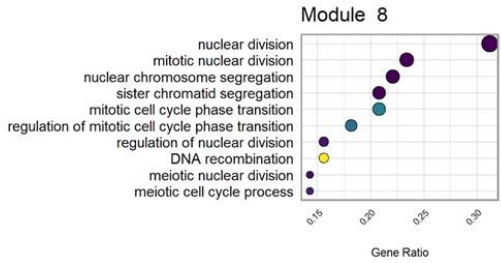
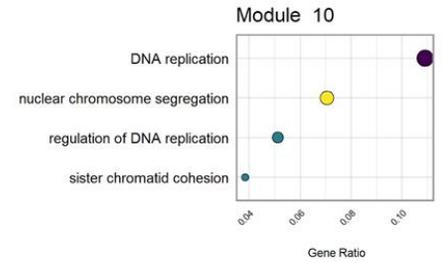
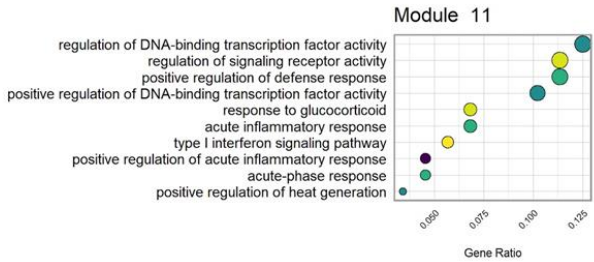
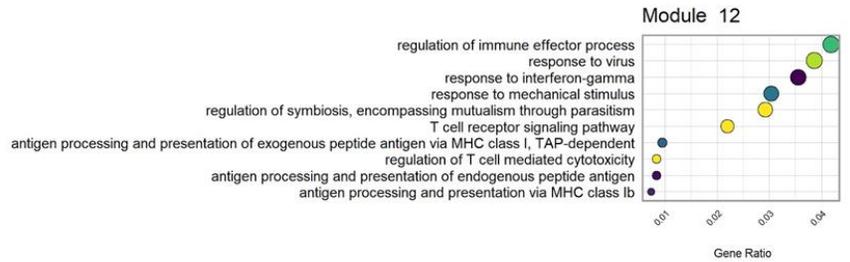
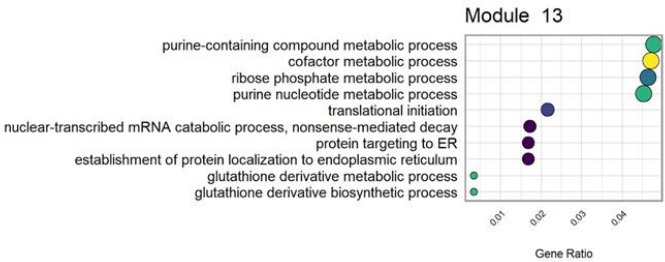
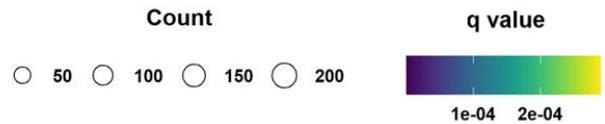

**Supplementary Figure 3.** Top 10 simplified enriched Gene Ontology terms of biological process in each of the relevant modules for coronaviruses infection. Go terms are ordered by q-value. Some modules are missing because they were not significantly enriched with any GO term.

### a

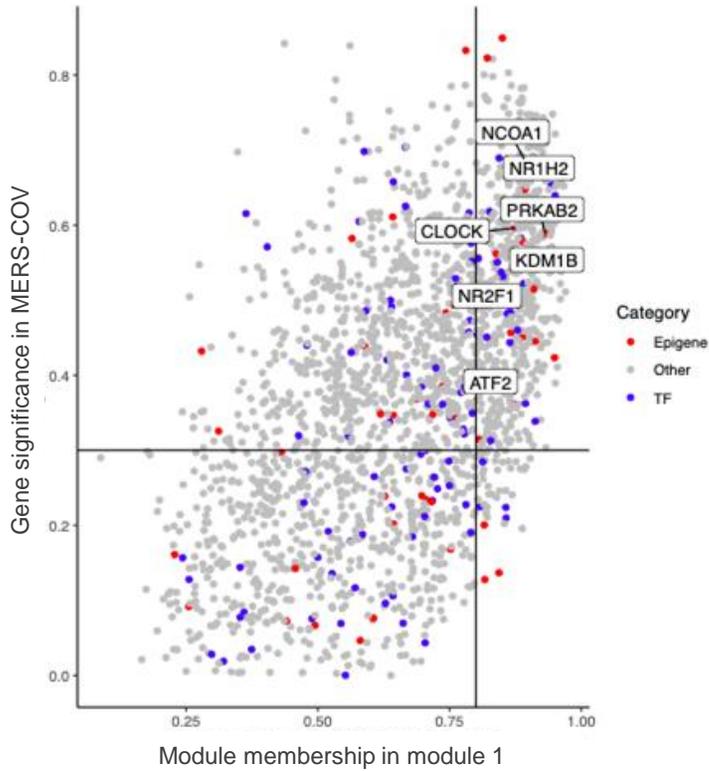

### b

| Gene | Function | Drug |
|---|---|---|
| NCOA1 | Nuclear receptor coactivator / H3 and H4 HAT | Etoposide, Tamoxifen |
| NR1H2 | Nuclear receptor / TF | Bexarotene, Diacerein, Alitretinoin, Isopropyl alcohol |
| PRKAB2 | Regulatory subunit of AMPK | Metformin, Fostamatinib Acetylsalicylic acid |
| CLOCK | Regulation of circadian rhythms (TF) | Salbutamol |
| KDM1B | Histone lysine demethylase | Tranylcypromine |
| ATF2 | TF / H2B and H4 HAT | Pseudoephedrine |

### c

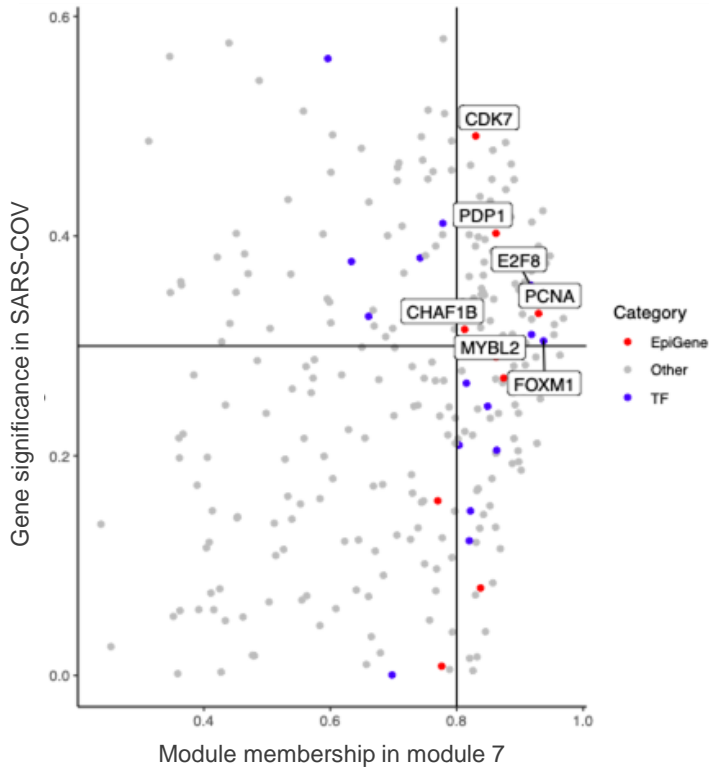

### d

| Gene | Function | Drug |
|---|---|---|
| CDK7 | Regulator of cell cycle progression / TF | Nintedanib, Erlotinib, Dasatinib, Crizotinib, Sorafenib, Ruxolitinib, Tofacitinib, Palbociclib, Masitinib, Midostaurin, Sunitinib, Neratinib, Imatinib, Lapatinib, Afatinib, Pazopanib, Bosutinib, Axitinib, Vandetanib, Gefitinib, Nilotinib |
| PCNA | Cofactor of DNA polymerase delta | Liothyronine Acetylsalicylic acid |

**Supplementary Figure 4. Potential epigenetic therapeutic targets for MERS and SARS-CoV. a** Module membership and Gene Significance of gene members of module 1. **b** Approved drugs for module 1's important epifactors and TFs. **c** Module membership and Gene Significance of gene members of module 7. **d** Approved drugs for module 7's important epifactors and TFs.